\documentclass[aps,twocolumn]{revtex4}

\usepackage{amsmath}
\usepackage{amssymb}
\usepackage[final]{graphicx}

\newcommand{\abs}[1]{\left|{#1}\right|}

\newcommand{\expect}[1]{\langle{#1}\rangle}

\newcommand{\sub}[1]{{\mbox{\scriptsize #1}}} 

\newcommand{\etaeff}{\eta_\sub{eff}}

\begin{document}
\title{Self-organization of atoms in a cavity field: 
threshold, bistability and scaling laws}
\author{J.\ K.\ Asb\'oth$^{1,2}$}
\author{P. Domokos$^{1}$}
\author{H. Ritsch$^{2}$}
\author{A. Vukics$^{1}$}

\affiliation{$^1$ Research Institute of Solid State Physics and Optics,
Hungarian Academy of Sciences\\
H-1525 Budapest P.O. Box 49, Hungary\\
$^2$ Institute of Theoretical Physics, University of Innsbruck, 
Technikerstrasse 25, A-6020 Innsbruck, Austria
}

\date{\today}

\begin{abstract}
  We present a detailed study of the spatial self-organization of
  laser-driven atoms in an optical cavity, an effect predicted on the
  basis of numerical simulations [P. Domokos and H.  Ritsch, Phys.\ 
  Rev.\ Lett.\ 89, 253003 (2002)] and observed experimentally [A. T.
  Black et al. in Phys.\ Rev.\ Lett.\ 91, 203001 (2003)]. Above a
  threshold in the driving laser intensity, from a uniform
  distribution the atoms evolve into one of two stable patterns that
  produce superradiant scattering into the cavity.  We derive
  analytic formulas for the threshold and critical exponent of this
  phase transition from a mean-field approach.  Numerical simulations
  of the microscopic dynamics reveal that, on laboratory timescale, a
  hysteresis masks the mean-field behaviour.  Simple physical
  arguments explain this phenomenon and provide analytical expressions
  for the observable threshold. Above a certain density of the atoms a
  limited number of ``defects'' appear in the organized phase, and
  influence the statistical properties of the system. The scaling of
  the cavity cooling mechanism and the phase space density with the
  atom number is also studied.
\end{abstract}

\maketitle

\section{Introduction}

The manipulation of cold atoms and molecules by laser light is a
rapidly growing field and has become a suitable ground for studying
fundamental phenomena of physics both experimentally and theoretically
\cite{metcalf}. In the last decade, the emphasis has partly been
shifted towards many-body effects in the dynamics of weakly
interacting atoms \cite{meystre,castin}.  The mechanical action of the
electromagnetic radiation field on free atoms rarely manifests these
effects. The refractive index of a cloud of atoms is simply composed
of the product of the single atom polarizability and the optical
density. Standard laser cooling methods were also conceived on the
basis of single-atom processes.  Only at densities as high as the ones
achievable in a magneto-optical trap does the dipole-dipole
interaction between atom pairs give rise to a Lorentz--Lorenz-type
refractive index and present an appreciable nonlinearity in the
optical density \cite{morice95,lagendijk97}. The underlying process,
the reabsorption of spontaneously scattered photons in the atomic
cloud \cite{labeyrie03a,labeyrie03b}, heats the atomic motion and
hence limits the attainable minimum temperature. In addition, this
effect also introduces spatial instability into the atomic cloud and
thus hinders degeneracy in phase space by optical means
\cite{sesko91}.

The mechanical effect of light on atoms inside a high-finesse
resonator is substantially modified with respect to free space, which
is the source of a variety of interesting phenomena in optical cavity
quantum electrodynamics \cite{hood00,pinkse00}.  The basic reason is
that a cavity photon makes many round-trips between the mirrors and
thus the back action of the atom on the field, enhanced by the cavity
finesse, cannot be neglected. As opposed to the external forces
exerted by laser fields, the light forces in a cavity cannot be
separated from the dynamics of the resonator mode, which is strongly 
influenced by that of the atom.

The coupled atom-field dynamics can yield an efficient damping of the
atomic motion via the mirror loss dissipation channel
\cite{horak97,domokos03}. Such ``cavity cooling'' schemes have
recently been demonstrated experimentally \cite{maunz04,nussmann05}.
The fact that cavity cooling allows for replacing the spontaneous
emission, which is the dissipation channel in all laser cooling
schemes, by irreversible photon loss from the cavity has important
consequences.  First, the internal structure of the atom is not
important and the mechanism can be operated on a wide range of
species. Second, the problem of the reabsorption of spontaneously
scattered photons, the source of the instability of atomic clouds at
high densities, can be suppressed.

The dynamics of atoms in a resonator is inherently a many-body problem
even at a small density of the ensemble \cite{munstermann00,black05}.
As all atoms are coupled to the same cavity mode, the modification of
the field by one atom is experienced by a remote atom as well as by
itself.  The cavity cooling mechanism may become inefficient since the
delicate dynamical correlation between one atom and the field mode
could be perturbed by the motion of another atom \cite{fischer01,asboth04}.
Indeed, one of the cavity cooling schemes was found to slow down
linearly with increasing number of atoms \cite{horak01}.

In a recent Letter we have predicted a cooperative behavior of the
atoms driven by a laser in a direction perpendicular to the axis of a
standing wave cavity \cite{domokos02b}. At high pump laser intensities
(above a threshold) the homogeneous atomic cloud self-organizes into
one of two regular checkerboard patterns that maximize scattering into
the cavity.  The constructive interference of fields radiated by the
individual atoms produces an intensity which depends quadratically on
the number of atoms (superradiance).  Corresponding to the two
patterns, there are two possible phases of the output field with 180
degrees difference, which have been observed in an experiment by
Black, Chan, and Vuletic \cite{black03}.

The onset of self-organization is relatively fast, on the microsecond
time scale. A basic property of the present system is that the field
created by the atoms traps and simultaneously cools them so that the
organized pattern remains stable on a long time scale (10's of ms).
The cavity cooling mechanism now acts on many atoms without losing
efficiency. There is no external finite-temperature heat bath to
define the temperature which, instead, is set by the dynamical
equilibrium of the dipole force fluctuations and the cavity cooling
effect. This is a distinctive feature with respect to the recently
demonstrated collective atomic recoil laser in a ring cavity (CARL)
\cite{kruse03,nagorny03,elsasser04,slama04}, where a magneto-optical trap is necessary to stabilize
the organized phase and the otherwise transient gain
\cite{bonifacio94}, and also to inject noise for obtaining the
phase transition-like behavior \cite{javaloyes04,cube04,robb04}.

In the present paper we discuss in detail the self-organization
process, from the viewpoint of phase transitions. A mean-field
approach leads to a well-defined threshold in the pumping strength.
Comparison to numerical simulations reveals effects scaling unusually
with the atom density, a characteristic feature of this cavity-coupled
many-atom system.

The paper is organized as follows. In Sect.\ \ref{sec:model}, the
equations of a semiclassical model are recapitulated, where the atoms
are represented as simple linearly polarizable particles.  Thereby the
theory applies to a much wider class of particles than alkali atoms.
The main features of the self-organization process, such as time
scales, superradiance, collective cooling, are surveyed using a
numerical example in Sect.\ \ref{sec:selforg}. Then, in Sect.\ 
\ref{sec:mean-field}, we introduce a one-dimensional mean-field model
and determine the threshold and critical exponent.  In Sect.\ 
\ref{sec:beyond}, we present the results of detailed numerical
simulations, which show effects beyond mean-field.  The atom number
enters the physics of the system in a form other than the density.
Above a certain atom number, stable defects appear in the self-organized 
pattern and modify the system
properties, which is accounted for in Sect.\ \ref{sec:defect}. In
Sect.\ \ref{sec:collective}, the cooperative atomic behavior is
discussed in detail by demonstrating the superradiance, and the
ensuing improvement of localization by collectivity. We conclude in
Sect.\ \ref{sect:conclusion}.

\section{Semiclassical model}
\label{sec:model}

\begin{figure}[htbp]
  \centering
  \includegraphics[width=7.5cm]{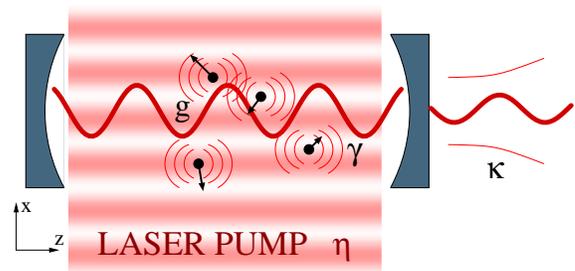}
  \caption{The setup: transversely driven atoms moving in a 
    standing-wave cavity. The laser intensity is given by the maximum
    free-space Rabi frequency (pumping strength) $\eta$.  The atoms
    couple to the cavity mode with 1-photon Rabi frequency $g$. The
    loss channels are: spontaneous emission (rate $2 \gamma$) and
    cavity loss (rate $2\kappa$).}
  \label{fig:scheme}
\end{figure}
We consider $N$ atoms in an open optical resonator (Fig.\ 
\ref{fig:scheme}).  The atoms are illuminated from the side by a
``transverse'' standing-wave laser pump with frequency $\omega$. This
geometry corresponds to various experimental setups realizing the
controlled transport of atoms from the side into a cavity
\cite{sauer04,nussmann05} where the standing-wave pump amounts to a
``conveyor belt'' \cite{schrader01,schrader04}.  There is an efficient
scattering of photons into the cavity mode quasi-resonant with the
pump, $\omega_C \approx \omega$, due to the enhanced dipole coupling
described by the single-photon Rabi frequency $g=\omega_C^{1/2}[2
\epsilon_0 \hbar V]^{-1/2} d_{eg}$, for a mode volume $V$ and atomic
transition dipole moment $d_{eg}$ along the cavity mode polarization.
Large detuning of the laser from the atomic transition
$|\omega-\omega_A| \gg \gamma$, where $2 \gamma$ is the full atomic
linewidth at half maximum, ensures that the upper level of the atoms
can be adiabatically eliminated. This model then describes a very
general class of linearly polarizable particles---in the following, we
continue to use ``atoms'' for convenience.  For the sake of
simplicity, we restrict the atomic motion to two dimensions, along the
pump laser and the cavity axis, coordinates $x$ and $z$, respectively,
without losing any relevant physical effect.  Motion in the third
dimension could be taken into account in the same way as along $x$.

The quantum master equation for the density matrix reads 
\begin{equation}
\label{eq:ME}
  \dot{\rho} = -\frac{i}{\hbar}\bigl[H,\rho\bigr] + {\cal L}\rho\; .
\end{equation}
Here the Hamiltonian is 
\begin{subequations}
\begin{equation}
\label{eq:H}
H = \sum_{j=1}^N \frac{{\bf p}_j^2}{2 M} - 
 \hbar \Delta_C a^\dagger a +
  \hbar U_0 \sum_{j=1}^N E^\dagger({\bf r}_j) E({\bf r}_j)
 \; ,
\end{equation}
where $a,a^\dagger$ are the boson operators of the cavity mode, ${\bf
  r}_j =(x_j, z_j)$ and ${\bf p}_j = ({p_x}_j, {p_z}_j)$ are the
position and momentum vectors of the $j$th atom. The Liouville
operator describing the cavity photon losses with rate $2\kappa$, and
the spontaneous emission reads
\begin{align}
 \label{eq:L}
 & {\cal L} \rho = 2\kappa \left( a \rho a^\dagger 
   - \frac{1}{2} a^\dagger a \rho - \frac{1}{2} \rho a^\dagger a 
 \right)
 \nonumber\\
 & -\Gamma_0 \sum_{j=1}^{N} \Bigl(E^\dagger({\bf r}_j) E({\bf
   r}_j) \rho + \rho E^\dagger({\bf r}_j) E({\bf r}_j) \nonumber\\
 & - 2 \int d^2{\bf u}\; N({\bf u}) E({\bf r}_j) e^{-i k_A {\bf u}
   {\bf r}_j} \rho e^{i k_A {\bf u} {\bf r}_j} E^\dagger({\bf
   r}_j)\Bigr)\; .
\end{align}
In the above formulas, $E({\bf r})$ is the dimensionless electric field, 
\begin{equation}
  \label{eq:Efield}
  E({\bf r}) =  f({\bf r}) a + \eta({\bf r})/g \approx \cos(k z)
 a + \cos(k x) \eta/g \; .
\end{equation}
\end{subequations}
The Rabi frequency of the driving laser is $\eta({\bf r})$, whose
position dependence is given by a $\cos(k z)$ mode function for a
standing-wave field. In the following, we are going to refer to the
maximum value of the Rabi frequency $\eta$ as ``pumping strength''.
The variation of the pump field along the cavity axis and that of the
cavity mode function $f(\bf r)$ along the transverse direction
(Gaussian envelope) are neglected.  The detunings are defined as
$\Delta_C=\omega-\omega_C$ and $\Delta_A=\omega-\omega_A$.  The
parameters
\begin{equation}
\label{eq:parameter}
U_0 = \frac{g^2 \Delta_A}{\Delta_A^2+\gamma^2} \; ,\;
\Gamma_0 = \frac{g^2 \gamma}{\Delta_A^2+\gamma^2}\; ,
\end{equation}
describe the dispersive and absorptive effects of the atoms,
respectively, as they shift and broaden the resonance line of the
cavity. In the last term of Eq.\ (\ref{eq:L}), the integral represents
the averaging over the angular distribution $N({\mathbf u})$ of the
random recoil due to spontaneous emission into the free-space modes.

Instead of directly using the density matrix, we consider the
evolution of the corresponding joint atom-field Wigner function
\cite{domokos01}.  This can be systematically approximated by
semiclassical equations for a set of classical stochastic variables,
$\alpha$, ${\bf p}_j$, ${\bf r}_j$, the index $j=1, \ldots, N$
labeling the atoms,
\begin{subequations}
\label{eq:sde}
\begin{multline}
\label{eq:sde_al}
  \dot\alpha = i \Big[\Delta_C - U_0 \sum_j \cos^2(k z_j)\Big] \alpha
  - \Big[ \kappa + \Gamma_0 \sum_j \cos^2(k z_j)\Big] \alpha
  \\
  - \etaeff \sum_j \cos(k z_j) \cos(k x_j) + \xi_\alpha \,,
\end{multline}%
\begin{multline}
\label{eq:sde_px}
\dot {p_x}_j = - \hbar U_0 (\eta/g)^2 \frac{\partial}{\partial x_j}
\cos^2(k x_j)
\\
- i \hbar (\etaeff^* \alpha - \etaeff\alpha^*)
\frac{\partial}{\partial x_j} \cos(k x_j) \cos(k z_j)+ {\xi_x}_{j}\; ,
\end{multline}
\begin{multline}
\label{eq:sde_pz}
  \dot {p_z}_j = - \hbar U_0 |\alpha|^2 \frac{\partial}{\partial z_j}
  \cos^2(k z_j)
  \\
  - i \hbar (\etaeff^* \alpha - \etaeff\alpha^*)
  \frac{\partial}{\partial z_j} \cos(k x_j) \cos(k z_j) + {\xi_z}_{j}\; ,
\end{multline}
\end{subequations}
where the effective pumping strength for the cavity mode is
\begin{equation}
\label{eq:eff_pumping}
\etaeff = \frac{\eta g}{-i\Delta_A+\gamma}\, .
\end{equation}
These equations include Langevin noise terms $\xi_\alpha$, ${\xi_x}_j$,
and ${\xi_z}_j$, defined by the non-vanishing second-order correlations,
\begin{subequations}
 \label{eq:noise_sde}
\begin{align}
  \langle \xi_\alpha^* \xi_\alpha \rangle &= \kappa  + \sum_{j=1}^N
  \Gamma_0 \cos^2(k z_j) \;, \\ 
  \langle {\xi}_{n} \xi_\alpha \rangle &= i \hbar \Gamma_0 \partial_n{\cal
  E}({\bf r}_j) \cos(k z_j) \;,\\ 
  \langle \xi_{n} \xi_{m} \rangle &= 2 \hbar^2 k^2 \Gamma_0 |{\cal E}({\bf
  r}_j)|^2 \overline{ u_n^2} \delta_{nm} + \hbar^2 \Gamma_0  \nonumber \\ 
 &  \Bigl[\partial_n {\cal E}^*({\bf r}_j) \, \partial_m{\cal
  E}({\bf r}_j) + \partial_n {\cal E}({\bf r}_j) \, \partial_m{\cal
  E}^*({\bf r}_j) \Bigr] \; ,
\end{align}
\end{subequations}
where the indices $n,m=x_j, z_j$. The noise terms associated with
different atoms are not correlated. The complex dimensionless electric
field ${\cal E}({\bf r})$ is derived from Eq.\ (\ref{eq:Efield}), 
replacing the field operator $a$ by the complex variable $\alpha$. We
iterate the coupled, stochastic Ito-type differential equations
(\ref{eq:sde}) by a Monte Carlo-type algorithm.

There are two types of force terms in the equations of the momentum
components. The terms in the first lines derive from the usual
one-dimensional ``optical lattice'' potentials, the laser pump keeps
the atoms inside the resonator via this term. In the second lines, the
force terms originate from the coherent redistribution of photons
between the pump and the field mode. The potentials depending on the
amplitude $\alpha$, which itself is a variable, are not conservative
(all but the optical lattice created by the transverse pump). The
time-delayed correlations in the dynamics of the atomic motion and the
field mode can result in a friction force on the atoms, known as
cavity cooling
\cite{horak97,hechenblaikner98,doherty00,vuletic01,vanenk01,domokos01,murr03,domokos04}.

\section{Self-organization}
\label{sec:selforg}

\begin{figure}[htbp]
  \centering
  \includegraphics[width=8cm]{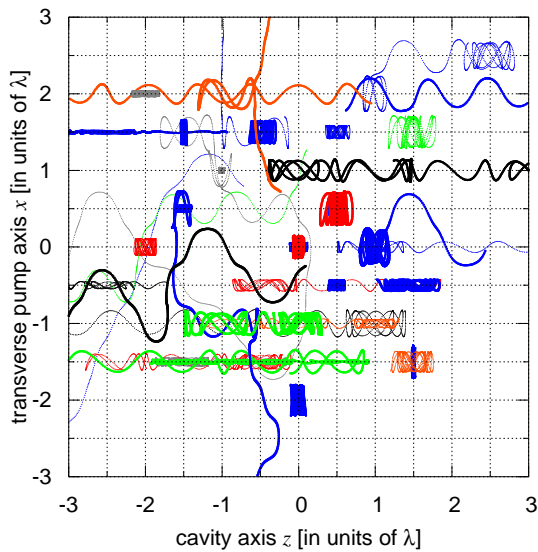}
  \caption{Two-dimensional trajectories of 40 rubidium atoms in a cavity, during 
    the initial 50 $\mu$sec. A checkerboard pattern of trapped atoms
    emerges, untrapped atoms move mainly along the cavity axis.
    Parameters: $\gamma=20/\mu$sec, $(g, \kappa)= (2.5,
    0.5)\gamma$, atomic detuning $\Delta_A =-500 \gamma$, cavity
    detuning $\Delta_C =- \kappa+N U_0$, and the pumping strength
    $\eta=50\gamma$.}
  \label{fig:selforg2D}
\end{figure}

We study he motion of the atoms in the cavity by numerically
integrating the set of stochastic ordinary differential equations
(\ref{eq:sde}). To be specific, ${}^{85}$Rb were considered, with the
$5^2\mathrm{S}_{1/2}, \mathrm{F}=3$ $\leftrightarrow$
$5^2\mathrm{P}_{3/2}, \mathrm{F}=4$ transition. Starting from a gas of
thermal atoms (random positions from a uniform, and velocities from a
thermal distribution) and no light in the cavity mode ($\alpha=0$),
with the right choice of parameters we observe a buildup of the cavity
field accompanied by the appearance of an organized pattern in the
spatial distribution of the atoms. This is illustrated in Fig.\ 
\ref{fig:selforg2D}, where the trajectories of 40 atoms during the
initial 50 \(\mu\)s of a run are shown.  The grid lines denote points
of maximum coupling to the standing-wave cavity or pump field.
Trapped atoms are oscillating about intersections of grid lines, where
a single atom can scatter pump photons into the cavity mode most
efficiently.  For many atoms, however, destructive interference can
inhibit the scattering process: the source term in Eq.\ 
(\ref{eq:sde_al}) contains the factor $\sum_j\cos(kx_j) \cos(kz_j)$,
which can be small even if all the atoms are maximally coupled due to
the alternating signs of the summands. In contrast to this, in Fig.\ 
\ref{fig:selforg2D}, only every second ``maximally coupled'' site---%
the black or the white fields of a checkerboard---is
occupied, leading to an efficient Bragg scattering of pump photons
into the cavity.

The emergence of a checkerboard pattern of atoms with every second
point of maximum coupling empty happens only due to the good choice of
the parameters ensuring positive feedback, as explained in the
following. Initially, in the random position distribution some
atoms scatter into the cavity in a given phase and some with an
opposite phase, and thus most of the scattered field is canceled.  The
dipole force (first term of Eq.\ (\ref{eq:sde_px})) attracts atoms
towards antinodes of the pump (for red detuning, $\Delta_A < 0$), but
almost no field in the cavity means no substantial modification of the
uniform position distribution along the cavity axis. This can be seen
in Fig.~\ref{fig:selforg2D}, where most atoms are well trapped along
the transverse axis but some meander along the cavity axis.  Due to
statistical fluctuations either the in-phase or opposite-phase
scatterers will be in tiny majority, and a small cavity field does
build up.  The dipole force due to the cavity field, first term of
Eq.\ (\ref{eq:sde_pz}), now attracts atoms towards antinodes of the
cavity.  The crucial point to consider is the interference of the
cavity and pump fields, giving the second terms in Eqs.\ 
(\ref{eq:sde}b,c). The product $\cos(kx_j)\cos(kz_j)$ alternates sign
between the black and white fields of the checkerboard. For a right
choice of detuning $\Delta_C$, there is a positive feedback and the
atoms are attracted towards the ``majority'' sites and are repelled
from the ``minority'' sites, due to the interference.

\begin{figure}[htbp]
  \centering
  \includegraphics[width=8cm]{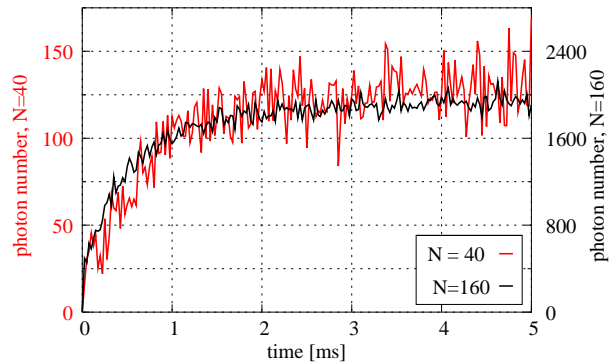}
  \caption{The time evolution of the photon number in the cavity on a
    long time scale, for $N=40$ and $N=160$ atoms (note the different 
    vertical scaling). 
    The parameters are
    the same as in Fig.~\ref{fig:selforg2D}}  
  \label{fig:slowselforg_a}
\end{figure}

The initial fast buildup continues over a longer timescale, with the
kinetic energy of the oscillating and the free-flying untrapped atoms
dissipated owing to the cavity cooling mechanism (for the transverse
pumping case and for the chosen detuning $\Delta_C$, it is the one
described in Ref.\ \cite{domokos04}).  This leads to an increase of
the ratio of trapped atoms and to a stronger localization in the
vicinity of the antinodes.  Simultaneously, the coherent scattering
into the cavity improves, giving a slow increase in the cavity field
intensity. The time evolution of the photon number is plotted in Fig.\ 
\ref{fig:slowselforg_a} for this self-organization process of 40 and
160 atoms. In the latter case, the photon number scale was rescaled by
a factor of 16. This way, the overlap of the two curves demonstrates
the superradiance effect, i.\ e., the intensity is quadratically
proportional to the atom number.

In this phase transition-like process, the reduction of the kinetic
energy is not a good characterization of the cooling efficiency.  For
the motion along the cavity axis, the transition from vanishing photon
number to the ``superradiated'' light field is accompanied by a change
of the heat capacity of the atomic ensemble since the number of
quadratic degrees of freedom changes from one (kinetic energy) to two
(kinetic and potential energy). In deep harmonic traps, even a very
low level of excitation can correspond to high kinetic energies. Thus
even though the temperature can increase, cavity-induced dissipation
increases the phase space density of the system by improving
localization. An appropriate measure of this process is the effective
phase space volume of the system (the inverse of the phase space
density), measured by the Heisenberg uncertainty product $\Delta
x\Delta p_x/\hbar$ and $\Delta z\Delta p_z/\hbar$ for each degree of
freedom.  Here $(\Delta x)^2$ and $(\Delta z)^2$ are the averages of
the squared distance of the atom from the nearest antinode along the
direction $x$ and $z$, respectively. For a harmonic potential, the
dimensionless effective phase space volume amounts precisely to the
mean number of excitation quanta. For untrapped atoms with mean
kinetic energy $\expect{p^2/m} = k_B T$, the phase space volume is
$\lambda \sqrt{m k_B T}/(4\sqrt{3}\hbar)$.

\begin{figure}[htbp]
  \centering
  \includegraphics[angle=270,width=8cm]{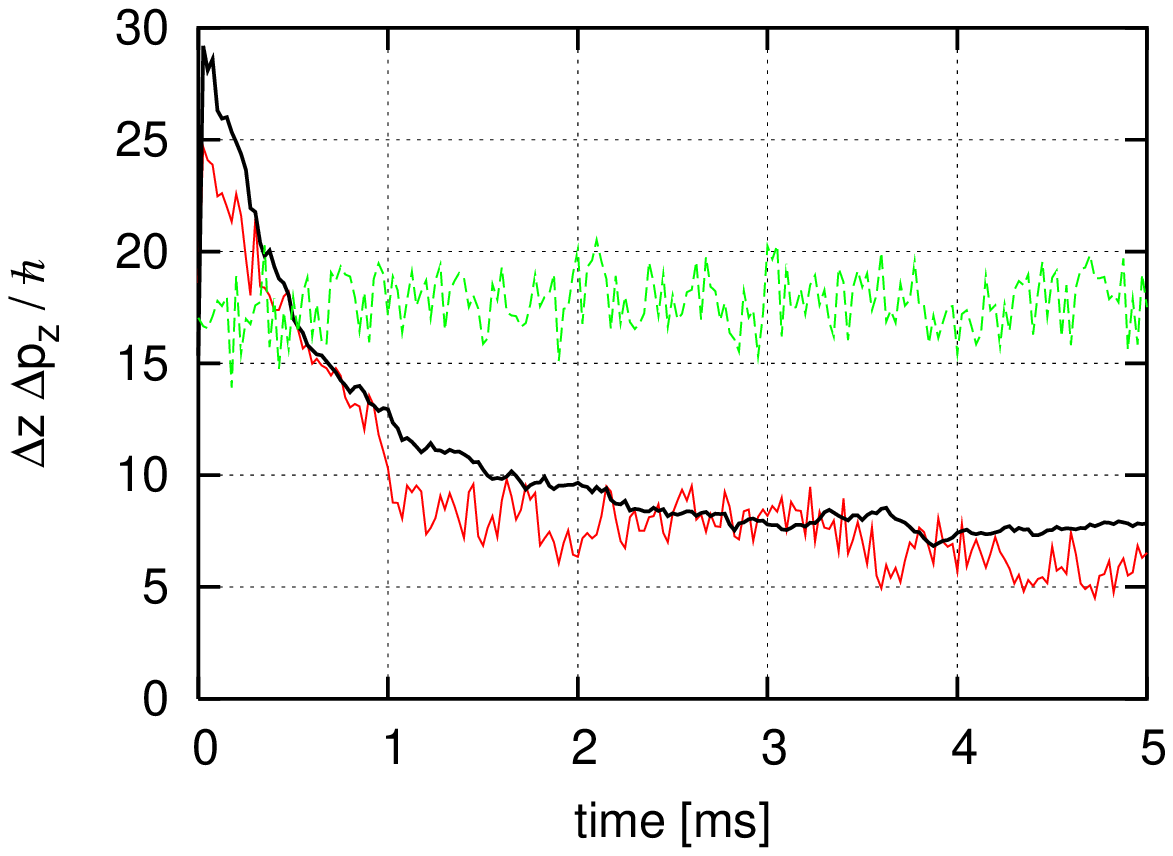}
   \includegraphics[angle=270,width=8cm]{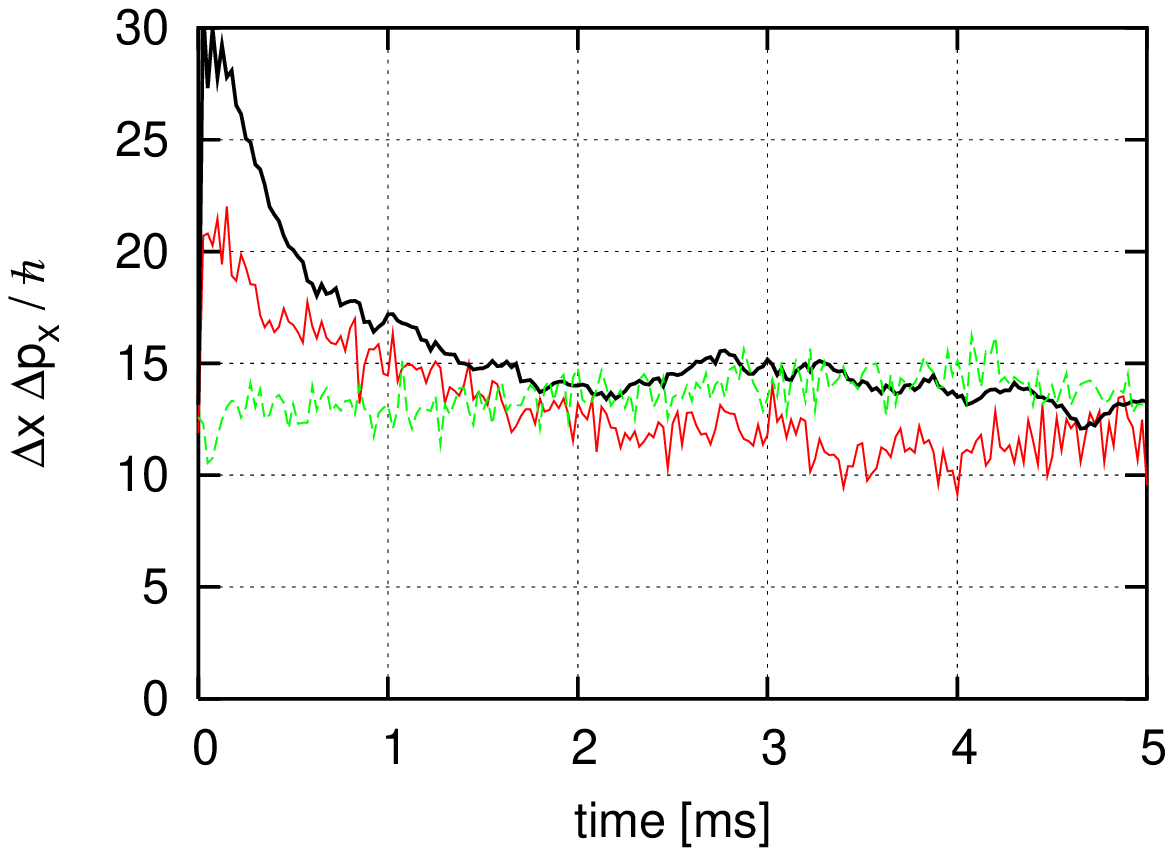}
  \caption{Evolution of the phase space volume, along the cavity 
    axis (up) and along the transverse pump (bottom), on a long time
    scale for various settings, same parameters as in Fig.\ 
    \ref{fig:slowselforg_a}. The number of atoms is N=40 and N=160
    (black line), and N=40 but pumped below threshold,
    $\eta=10\gamma$, for the dashed line.}
  \label{fig:slowselforg_b}
\end{figure}

In Fig.\ \ref{fig:slowselforg_b}, the time evolution of the Heisenberg
uncertainty product is shown for three cases.  When the pumping
strength is below threshold ($N=40$, \(\eta=10 \gamma\)), the
uncertainty product is a constant. Here the spatial distribution of
the atomic ensemble remains uniform---the transverse pump is too weak
to induce any noticeable spatial modulation at temperatures $k_B T
\approx \hbar \kappa$.  Thus $\expect{\Delta z \Delta
  p_z}=\textrm{const.}$ and $\expect{\Delta x \Delta
  p_x}=\textrm{const.}$ reveal that the temperature itself does not
change in either direction. At the cavity cooling limit $k_B T = \hbar
\kappa$ the numerical value for the phase space volume from the last
paragraph is $13.4 \hbar$, in accordance with the value along the
transverse $x$ direction. Along the cavity axis $z$, however, the
phase space volume is above this estimate, indicating a higher
temperature.  Above threshold, the phase space volume transiently
jumps to high values for both directions, and then it is gradually
decreased.  Compression is apparently more efficient along the cavity
axis, here the phase space goes considerably below the value
corresponding to the uniform distribution. In both plots the two
curves corresponding to $N=40$ and $N=160$ are very similar, which
manifests that the cooling rate of the ensemble is independent of the
atom number.  This is a very important observation, being at variance
with the expectation that the efficiency of cavity cooling mechanism
is reduced for increasing number of atoms. This prediction was made
for a setup where the external pump field is injected directly into
the cavity.  Apparently it does not apply to the transverse pumping
case studied here.

With the help of these few examples, we surveyed three important
properties appearing in the dynamics of a laser-driven atomic ensemble
coupled to a cavity mode: (i) the system rapidly self-organizes into a
checkerboard pattern in a trapping field, (ii) which is generated by a
collective, superradiant scattering into the cavity, and finally,
(iii) the energy of the atoms is dissipated at a rate independent of
the number of atoms. This behaviour requires a
sufficiently strong pumping strength, indicating the possibility of a
well-defined threshold separating two different stability regions.
This threshold is discussed in the next section within the framework
of a mean field approximation.

\section{Mean-field approximation}
\label{sec:mean-field}

The essence of the self-organization process can be understood on the
ground of conservative mean-field forces acting on the atoms. This
amounts to treating the cavity field as if it responded immediately to
the positions of the atoms. Cavity cooling, which is directly related
to the time lag of the cavity field, is absent in this model.
Moreover, the mean-field approach corresponds to the thermodynamic
limit of the system: $N\to\infty$, $g\to 0$, $\kappa=\textrm{const}.$
with $Ng^2=\textrm{const}.$ Physically, the limit can be thought of as
taking larger cavities (cavity length $l_{\rm cav}\to \infty$) filled
with a gas of atoms of constant density (atom number $N\propto l_{\rm
  cav}$).  Due to the $V^{-1/2}$ dependence of the coupling constant
$g$ on the mode volume $V$, one then has $Ng^2=\textrm{const}.$,
neglecting the variation of the waist of the mode. Moreover, due to
larger photon travel time between the mirrors, the reflectivity has to
scale like $\propto l_{\rm cav}^{-1}$ to keep $\kappa=\textrm{const}.$
For the sake of simplicity, we analyze one-dimensional motion along
the cavity.

\subsection{Potentials}
Taking $x_j=0$, $j=1, \ldots, N$, according to Equation (\ref{eq:sde_pz}) 
each atom moves in a potential
\begin{equation}
\label{eq:potential}
V(z) = U_2 \cos^2 (k z) + U_1 \cos (kz)\,
\end{equation}
composed of the sum of a \(\lambda/2\) periodic potential stemming from 
the cavity field and a \(\lambda\) periodic
one arising from the interference between cavity and pump fields.
The potential depths are given by
\begin{subequations}
\begin{align}
\label{eq:prefactors}
U_2 &= N^2 \langle \cos(kz) \rangle^2 \, \hbar I_0\,U_0\\
U_1 &= 2 N  \langle \cos(kz) \rangle \, \hbar I_0 \, 
(\Delta_C - N U_0 \langle \cos^2(kz) \rangle)\;.
\end{align}
\end{subequations}
These, in the mean-field approximation, depend on the position of the
individual atoms only via the mean value
\begin{equation}
  \label{eq:order}
\Theta = \langle \cos(kz) \rangle = \frac{1}{N}\sum_{i=1}^{N} \cos(k z_i)  \; ,
\end{equation}
which can be considered a spatial order parameter, and via
the bunching parameter
\begin{equation}
  \label{eq:bunching}
  {\cal B} = \langle \cos^2(kz) \rangle = \frac{1}{N} \sum_{i=1}^{N} 
\cos^2(k z_i)  \; .
\end{equation}
The order parameter \(\Theta\) has characteristic values: (i) \(\Theta
\approx 0\) describes the uniform distribution, (ii) \(\Theta
\rightarrow \pm 1\) corresponds to a self-organized phase with atoms
in the even or odd antinodes, respectively.  Finally, $I_0$ represents 
the maximum number of photons each atom can scatter into the cavity:
\begin{equation}
  \label{eq:I_0}
  I_0 = \frac{\abs{\etaeff}^2}{[\kappa + N \Gamma_0 {\cal B}]^2+
[\Delta_C - N U_0 {\cal B}]^2} \; .
\end{equation}

In equilibrium the spatial distribution of the atoms and the above
averages are time-independent, which makes it possible to attribute a
physical meaning to the potential (\ref{eq:potential}).  Since the
potential depends on the position distribution, however, the system is
highly nonlinear.

For $U_0<0$, obviously $U_2<0$ and the cavity field gives a
potential with ``even'' wells at $kz = 2n\pi$ and ``odd'' ones at
$kz=(2n+1)\pi$.  The interference term ($\propto \cos(kz)$)
discriminates between the even and odd sites, raising the energy of
one of them and lowering that of the other. If $2\abs{U_2}<\abs{U_1}$
this effect is so strong that $V(z)$ yields a potential with wells at
the even and hills at the odd sites -- or the other way around,
depending on the sign of $U_1$.

The sign of $U_1$ is crucial. To simplify the dependence, let us
require a cavity detuning $\Delta_C<-N\abs{U_0}$ so that the second
factor in $U_1$ is always negative regardless the momentary
configuration of the atoms. To be specific, in the following we are
going to use
\begin{equation}
\label{eq:dc}
\Delta_C = N U_0 - \kappa\; .
\end{equation}
If the atoms accumulate around the even (odd) antinodes, then \(\Theta
\approx +1\) ( \(\Theta \approx -1\) ) and the $U_1 \cos{kz}$
potential is attractive at the even (odd), while repulsive at the odd
(even) sites. Therefore Eq.\ (\ref{eq:dc}) is the proper choice for
positive feedback that makes the runaway solution of self-organization
possible.

Two more parameters describing spatial order are used later in this
work: (i) the ``defect ratio'', the ratio of atoms closer to minority
sites than majority sites; (ii) the ``localization parameter'', the position
variance (along the cavity axis and/or the transverse pump)
\begin{equation}
  \label{eq:localization}
  {\cal D}_z = \frac{1}{N} \sum_{i=1}^{N} \left(\frac{k
  z_i}{\pi}\right)^2\; ,
\end{equation}
where $z_i$ is measured from the nearest antinode of the cavity mode
function.  A uniform distribution of atoms gives a defect ratio close
to 50 \% and a localization parameter of $1/12$.  For a self-organized pattern,
both parameters approach 0.

\subsection{Canonical distribution}

We suppose that the phase-space distribution of the atoms factorizes
to position and momentum dependence, the latter simply given by a
thermal distribution with mean energy $k_B T$. There is no external
finite-temperature heat bath to set $k_B T$, it is instead determined
by the dynamics (\ref{eq:sde}) through the equilibrium of the cavity
cooling and the Langevin noise terms. This allows for a position- and
time-dependent effective temperature, effects neglected in this model.
For a far-detuned pump, cavity losses dominate spontaneous emission,
and an estimate $k_B T \approx \hbar \kappa$ is provided by the
Einstein relation. The spatial density of the atoms in the potential
$V(z)$ is then given by a canonical distribution,
\begin{equation}
\label{eq:rho_mf}
\rho(z) = \frac{1}{Z} \exp(-V(z)/(k_B T))\; ,
\end{equation}
with the partition function $Z=\int \exp(-V(z)/(k_B T)) dz$ ensuring
that $\rho(z)$ is normalized to unity.  In our case the potential
\(V(z)\) is a function of the density \(\rho(z)\), therefore this
equation has to be solved in a self-consistent manner. 

\begin{figure}[htbp]
  \centering
  \includegraphics[width=8cm]{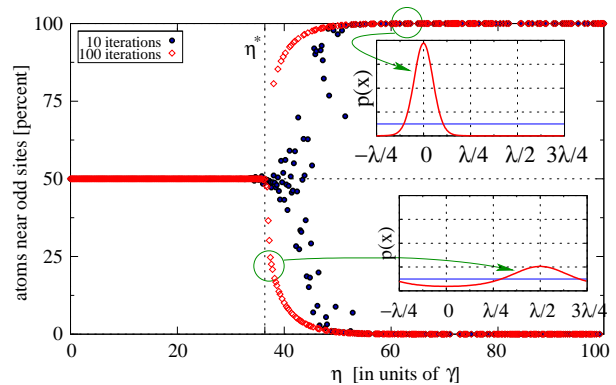}
  \caption{Numerical iterations of the mean-field approximation.
    The percentage of atoms near odd sites after 10 (full circles) 
    and 100 (empty diamonds) iterations is shown varying the pumping
    strength $\eta$.
    The vertical line is at $\eta^*$ of (\ref{eq:eta_crit}). The two
    insets show two steady-state position distribution functions at
    two different pumping strengths. 
    Parameters: $\kappa=\gamma/2, \Delta_A= -500 \gamma$, $N g^2 = 200
    \gamma^2$ $\Delta_C=-\kappa-Ng^2/|\Delta_A|$, $k_B T = \hbar \kappa$.}  
  \label{fig:ising}
\end{figure}

A direct approach to solving (\ref{eq:rho_mf}) is to use it
iteratively to determine the $\rho(z)$ for given values of the
physical parameters. 
We set the temperature to the cavity cooling limit $k_B T = \hbar
\kappa$. Note that the temperature parameter $T$ just rescales the
pumping strength $\eta$, e.g., taking $k_B T=2 \hbar \kappa$
would correspond to increasing $\eta$ by a factor of $\sqrt{2}$. The
results thus obtained after 100 iterations of (\ref{eq:rho_mf}) 
for an experimentally realistic example are
shown in Fig.~\ref{fig:ising} (empty diamonds).
There we plot the percentage of atoms near odd sites, i.e., with
$|(2m+1)\lambda/2-z|<\lambda/4$ for any integer $m$, as a function of
the pumping strength.  Below a certain threshold in the pumping laser
amplitude $\eta^*$ 
(vertical dotted line) the atoms are distributed uniformly, for
stronger pumping this symmetry is broken.  Two examples of such
self-organized position distributions obtained by the iterations are
shown in the insets.  Note also that the convergence of the iterations
is slow near the critical $\eta^*$ (critical slowing down), this is
evidenced by plotting the results after 10 iterations (full circles) 
as well.

The uniform distribution, $\rho(z)=1/\lambda$ leads to 
$\Theta=0$, ${\cal B}=1/2$, which give 
$V(z)=0$, thus the distribution fulfills (\ref{eq:rho_mf}) trivially for
any values of the physical parameters. To investigate its stability, we
add an infinitesimal perturbation to it:
\begin{equation}
  \label{eq:perturb}
  \rho^{(0)}(z)=\frac{1}{\lambda} + \epsilon g(z) \frac{1}{\lambda}\; ,
\end{equation}
with $\epsilon$ infinitesimal, and the general $\lambda$-periodic
perturbation function
\begin{align}
  \label{eq:perturb_func}
  g(z) &= \sum_{m=1}^\infty (A_m \cos(m kz) + B_m \sin(m kz)) \; ,\\
  & \sum_m A_m^2 + B_m^2 =1 \nonumber\; .
\end{align}
Since the spatial average of $g(z)$ disappears, $\rho^{(0)}(z)$ is
normalized to 1.
Iterating (\ref{eq:rho_mf}) once leads to the new density function
\begin{align}
\rho^{(1)}(z) &= 
\frac{1}{\lambda} - N \epsilon \frac{A_1}{\lambda} \frac{\hbar I_0}{k_B T} 
(\Delta_C - \frac{N}{2} U_0) \cos(kz) + o(\epsilon^2).
\end{align}
To lowest order in $\epsilon$ the only relevant perturbation is that
proportional to $\cos{kz}$. Stability requires that the first-order
correction in $\epsilon$ be self-contracting.
Substituting (\ref{eq:dc}) for the cavity detuning, we have the
following instability threshold for the uniform distribution:
\begin{align}
N I_0 \hbar \left(\frac{N}{2}\abs{U_0} +\kappa\right) > k_B T\; .
\end{align}
For far-detuned atoms $\abs{\Delta_A} \gg N g^2 / \kappa$, i.e., when the
total cavity mode shift by the atoms is much smaller than the cavity
linewidth $N \abs{U_0} \ll \kappa$, this translates to the following
threshold on the pumping strength:
\begin{align}
\label{eq:eta_crit}
\eta > \eta^* = \sqrt{\frac{k_B T}{\hbar \kappa}} 
\frac{\kappa \abs{\Delta_A}}{\sqrt{N} g} \sqrt{2}\; . 
\end{align}
This approximation of the critical value $\eta^*$ is in good
agreement with the simulations shown on Fig.~\ref{fig:ising}, giving
$\eta^*=35.4 \gamma$, which differs by less than $\gamma$ from the
actual value. To make the physical content more transparent, this
condition can be expressed in terms of the transverse pumping power
density (energy/unit area/unit time) as
\begin{equation}
  P_{\rm in} > k_B T \left(\frac{\Delta_A}{\gamma}\right)^2 \kappa 
\frac{4 k^3}{3 N/V}\; .
\end{equation}
As shown in the next Section, the condition (\ref{eq:eta_crit}) amounts 
to requiring that the potential depth along the cavity axis of a 
self-organized checkerboard pattern exceed significantly the energy
scale of thermal fluctuations.  

\subsection{Critical exponent}

The stability analysis of the uniform distribution of atoms 
$\rho(z)=1/\lambda$ has revealed the critical value of the pumping
strength $\eta^*$. Moreover, we have seen that for $\eta\approx\eta^*$, 
the relevant perturbation gives 
\begin{equation}
\label{eq:rho_attempt}
\rho(z)=1/\lambda + \epsilon \cos(kz)/\lambda.
\end{equation}
Going beyond first-order perturbations, the above formula 
allows us to solve Eq.~(\ref{eq:rho_mf}) self-consistently. 

Pumping the atoms slightly above threshold, $\eta=\eta^* (1+\delta)$,
with $\delta \ll 1$, we expect (\ref{eq:rho_attempt}) to give a good
approximation of $\rho(z)$, with $\epsilon$ depending nonlinearly on
$\delta$. Substituting it into (\ref{eq:rho_mf}) we need to expand the
exponential to third order to obtain a solution to lowest order in the
small parameters.  This analytical calculation yields $\epsilon
\propto \delta^{1/2}$, i.e., above the critical value $\epsilon$
increases from 0 as the square root of the dimensionless excess
pumping strength. The order parameter $\theta=\epsilon/2$ and the
percentage of majority atoms ($2\epsilon/\pi$) both depend linearly on
the small parameter $\epsilon$.  Therefore, the analytical result
shows the critical exponent of this phase transition to be
$\frac{1}{2}$.

\section{Numerical simulations of the phase transition}
\label{sec:beyond}

\begin{figure}[htbp]
  \centering
  \includegraphics[width=8cm]{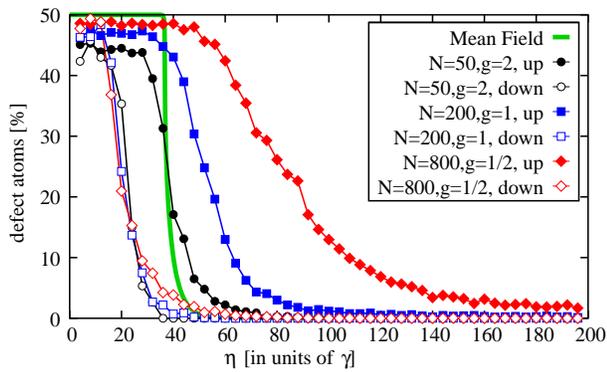}
  \caption{Ratio of defect atoms against pumping strength $\eta$, 
    4ms after the loading of the trap with a uniform (``up'') or
    organized (``down'') gas of atoms. The different curves show the
    approach towards the thermodynamic limit. The parameters are the
    same as in Fig.~\ref{fig:ising}.}
  \label{fig:threshold_Ng2}
\end{figure}

In the mean-field description of the steady state, the number of atoms
enters only in the form of the atomic density $N/V \propto N g^2$.
The approximation is expected to be valid in the thermodynamic limit,
i.e., $N\to \infty$ with the atomic density and the cavity loss rate constant.
Trying to approach this limit in simulations of Eq.~(\ref{eq:sde}) we
are in for a surprise.  In Fig.~\ref{fig:threshold_Ng2}, we show the
measured percentage of defect atoms after $4$ ms of simulation time as
a function of the pumping laser strength. The thermodynamic limit is
approached in three steps, $N=50, 200$, and $800$.  The parameters are
the same as in Fig.~\ref{fig:ising}, the atoms had random initial
velocities from a thermal distribution with average kinetic energy
$\hbar \kappa$. The initial positions were either uniformly
distributed (``up'') or at ``odd'' points of maximal coupling
(``down''). In this controlled way we mimic the ramping of the laser
power.  Although the ``down'' curves show reasonable agreement with
the mean field result, the ``up'' curves scale anomalously: a
hysteresis is observed, whose breadth increases with the atom number.

The hysteresis effect observed in Fig.~\ref{fig:threshold_Ng2} but
absent from the mean-field prediction is due to the long timescales
associated with reaching a steady state. In fact, thermal fluctuations
do not only alter the equilibrium solution by smoothing out the
concentration differences due to the dynamics (this is correctly
rendered by the mean field approach), but they can also delay
significantly the onset of that equilibrium. This slowing down is
effective if the energy scale of thermal fluctuations is at least of
the same order as the potential diffences due to the statistical
fluctuations in the initial positions of the atoms. 

The statistical fluctuations for a finite uniformly distributed atomic 
ensemble lead to $(N-\delta N)/2$ atoms around the $kz=2n
\pi$ and $(N+\delta N)/2$ atoms are around the $kz = (2n+1) \pi$
positions.  Taking uniform distributions around the respective
antinodes, we have $\sum \cos(k z_i) \approx 2 \delta N/\pi$ and $\sum
\cos^2(k z_i) \approx N/2$.  The potential difference then reads
\begin{align}
\label{eq:sp_pot_diff}
\Delta E &= 2\abs{U_1} = \hbar I_0
\frac{8\delta{N}}{\pi} (\kappa-N |U_0|/2)\nonumber\\
&= \hbar \frac{4\delta{N}}{\pi} \frac{\eta^2 g^2}{\kappa
  \Delta_A^2}\;.
\end{align}
For the final expression in the second line, we considered the
far-detuned regime, $\abs{\Delta_A} \gg \gamma \sim\kappa\sim g$,
and $N |U_0|, N \Gamma_0 \ll \kappa$. 

The self-organization occurs ``instantly'' if the trap depth
originating from the statistical fluctuations exceeds the average
kinetic energy $k_B T$ of the atoms. Using the expectation value of
the finite-size fluctuations, \(\delta N \approx \sqrt N\), we obtain
\begin{equation}
 \label{eq:threshold_cond}
\eta  > \eta_\uparrow = \sqrt{\frac{k_B T}{\hbar \kappa}}
\frac{\kappa\abs{\Delta_A}}{N^{1/4} g}\frac{\sqrt\pi}{2} .
\end{equation}
Comparison with (\ref{eq:eta_crit}) gives $\eta_\uparrow =
\sqrt{\pi/8}\; N^{1/4}\; \eta^*$: the functional dependence of the two
thresholds on the physical parameters are the same, except for the
``anomalous'' scaling of $\eta_\uparrow$ with $N$ as the thermodynamic
limit is approached.

\begin{figure}[htbp]
  \centering
  \includegraphics[width=8cm]{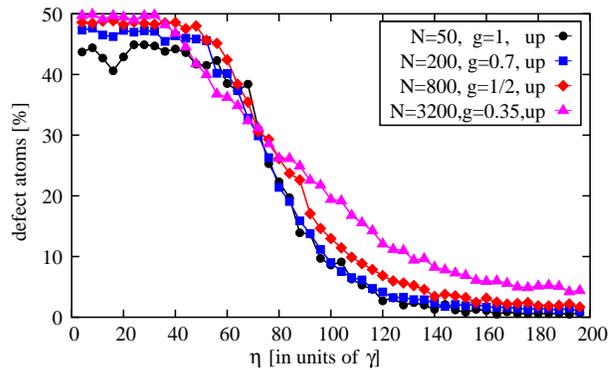}
  \caption{Ratio of defect atoms in the ensemble (measure of the spatial 
    order) as a function of the control parameter \(\eta\) (given in
    units of \(\gamma\)). There is a clear threshold for
    self-organization which is independent of the atom number provided
    \(Ng^4\) is kept constant. Parameters are the same as in
    Figs.~\ref{fig:threshold_Ng2} and \ref{fig:ising}.
  }
  \label{fig:threshold_Ng4}
\end{figure}

To check the laboratory timescale threshold for self-organization
(\ref{eq:threshold_cond}), we performed simulations with the same
physical parameters as in Figs.~\ref{fig:ising} and
\ref{fig:threshold_Ng2}, starting from uniformly distributed 
atoms, but this time increasing the atom number with
$Ng^4=\textrm{const}.$ In Fig.\ \ref{fig:threshold_Ng4}, the
percentage of defect atoms is plotted as a function of the pumping
\(\eta\).  These numerical results confirm that the threshold depends
on $N g^4$, moreover, the value 
$\eta_\uparrow = 83 \gamma$ from (\ref{eq:threshold_cond}) with $k_B
T=\hbar \kappa$ is also consistent with the simulations.  In the
transition regime, there is a remarkable overlap of the curves
corresponding to various number of atoms with keeping \(Ng^4\)
constant, indicating that the equilibration time also scales
with $Ng^4$.

The ``down'' curves of Fig.~\ref{fig:threshold_Ng2} are invariant with
respect to the thermodynamical limit, but the threshold they give is
only 50--80~\% of the mean-field prediction $\eta^*$ of Eq.~(\ref{eq:eta_crit}).  
Again, this can be explained with an argument
based on the comparison of energy scales. Instead of the statistical
fluctuations in the position distribution, we now have a well defined
initial state: $\sum \cos kz_i = \sum \cos^2 kz_i = N$.  The threshold
$\Delta E > k_B T$ now gives the particularly simple result: $\eta >
\eta_\downarrow=\eta^* / 2$. This is slightly below the ``down'' curves of 
Fig.~\ref{fig:threshold_Ng2}, the difference can be attributed to the 
nonoptimal coupling due to the position spread of the trapped atoms.  

In the last curve of Fig.~\ref{fig:threshold_Ng4}, $N=3200$, the
atom-cavity coupling $g$ is the smallest parameter, and thus the
system does not strictly realize the strong-coupling regime of cavity
QED. In fact, self-organization does not depend on strong coupling:
small $g$ can be compensated by increasing the number of atoms.  We
remark that the experimental setup of \cite{black03} was also operated
out of the strong coupling regime.

\section{Stable defect atoms}
\label{sec:defect}

The curve in Fig.~\ref{fig:threshold_Ng4} corresponding to $N=3200$
deviates slightly from the other data and for large pumping strength
it converges to a nonzero value of the defect atoms.  In fact, for the
parameters chosen there, defect atoms can be stably trapped at the
minority sites of the checkerboard of maximally coupled points, which
presents another important physical element of the system beyond the
capabilities of the simple mean-field theory.  In the following we
discuss the condition for the appearance of defects and the upper
limit on their number.

For the analytical treatment of the defect atoms we use the 1D model
of Section \ref{sec:mean-field}.  If the atoms are perfectly
self-organized, say, $kz_j=2n_j\pi$ with integer $n_j$ for every $j$,
then the potential depths of Eqs.\ (\ref{eq:prefactors}) scale with
the number of atoms as $U_2 = - N^2\, I_0 \abs{U_0}$, and $U_1 = - N
\, 2 I_0 \kappa$.  Thus, for large enough $N$, the $\lambda/2$-periodic
potential is dominant and atoms can be trapped in the minority sites.
These stable ``defect'' atoms reduce the superradiant scattering of
the self-organized pattern.

How many defects can stably reside in the pattern?  For simplicity, we
take $N-M$ atoms exactly at $kz=2n \pi$ and $M<N/2$ ``defect'' atoms
at $kz = (2n+1) \pi$, neglecting the position spread. We then
obviously have $\sum \cos(k z_i) = N-2M$ and $\sum \cos^2(k z_i) = N$.
Substituting this and the prescription (\ref{eq:dc}) for the cavity
detuning into (\ref{eq:prefactors}) we obtain
\begin{subequations}
\begin{align}
\label{eq:prefactors0}
U_2 &= - (N-2M)^2 \abs{U_0} \hbar I_0\;,\\
U_1 &= - 2 (N-2M) \hbar I_0 \kappa \;,\\
I_0 &= \frac{\abs{\etaeff}^2}{[\kappa+N\Gamma_0]^2+\kappa^2}\;.
\end{align}
\end{subequations}
Defect atoms can persist if at every $k z= n \pi$ there is a 
potential minimum: 
\begin{equation*}
0 < 2\abs{U_2}-\abs{U_1} = 2 \hbar I_0 (N-2M) 
\left[\left(N-2M\right)\abs{U_0} - \kappa\right]\;,
\end{equation*}
which entails
\begin{equation}
\frac{N}{2}-\frac{\kappa}{2\abs{U_0}} > M\;.
\end{equation}
If some defect atoms appear, the rise in their number is limited by
the above inequality. In particular, if the left-hand-side is
negative, there can be no stable defects: the condition for the
possibility of stable defects reads
\begin{align}
\label{eq:N_thr}
N > N_{\mathrm{thr}} &= \frac{\kappa}{\abs{U_0}}\;.
\end{align}
Note that this threshold is independent of the pumping strength.  The
maximum number of defects is limited by
\begin{equation}
  \label{eq:max_defect}
M < M_{\mathrm{max}} = \frac{N-N_{\mathrm{thr}}}{2}.  
\end{equation}
Working at large atomic detuning, $\abs{U_0} \approx
g^2/\abs{\Delta_A}$, we find that defects can appear if 
the total mode frequency shift due to the atoms exceeds the cavity
linewidth:
\begin{align}
\label{eq:density_defect}
N g^2 > \kappa \abs{\Delta_A}.
\end{align} 
This inequality puts a lower bound on the atomic density.
Equivalently, it amounts to an upper bound on the atomic detuning
$\Delta_A$: to avoid the occurrence of defects a large detuning can be
chosen. This, $\abs{\Delta_A} \gg N g^2/ \kappa$, is exactly the
``far-detuned'' limit of the previous Sections, used to derive the
thresholds $\eta^*, \eta_\uparrow$ and $\eta_\downarrow$. Likewise,
none of the curves in Fig.\ \ref{fig:threshold_Ng2} satisfy
Eq.~(\ref{eq:density_defect}).  On Fig.~\ref{fig:threshold_Ng4},
however, the curve corresponding to $N=3200$ is above the critical
density (\ref{eq:density_defect}).

For comparison to the full solution of the dynamics, we numerically
simulated the equations (\ref{eq:sde}) at fixed $\kappa=\gamma/2$,
$g=5\gamma/2$, $\Delta_A=-500 \gamma$, and $\eta=50 \gamma$. The
number of atoms was varied from 0 to 200, 25 runs with different
random seeds were performed for each atom number for a duration of 5
ms.  The conditions for self-organization derived in Sects.\ 
\ref{sec:mean-field} and \ref{sec:beyond}, give for this parameter
setting a threshold atom number $N \gtrsim 10$--$40$ (for $k_B T =
\hbar\kappa$--$\hbar \gamma$).

\begin{figure}[htbp]
  \centering
  \includegraphics[angle=270,width=8cm]{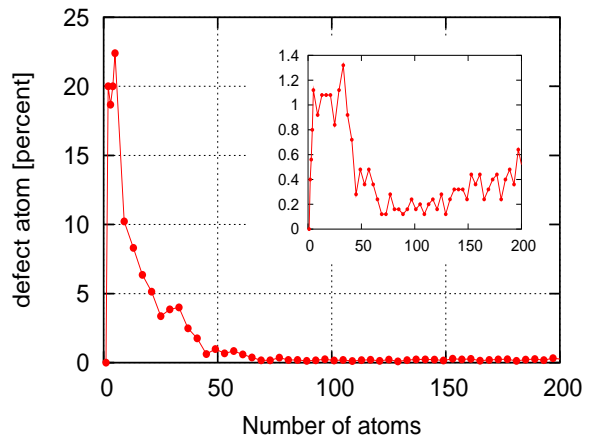}
  \caption{Defect atom ratio 5 ms after loading the trap 
    as a function of the atom number. Each point is an average over
    25 runs of the simulation. The inset shows the average number
    defects.  The physical parameters are:
    $\kappa=\gamma/2$, $g=2.5 \gamma$, $\Delta_A=-500 \gamma$,
    $\eta=50 \gamma$.}
  \label{fig:varn_defect}
\end{figure}
The numerical results presented in Fig.\ \ref{fig:varn_defect} show
that the ratio of defect atoms averaged over the 25 trajectories is
well below 50\%. For small atom numbers this is merely a ``finite
size'' effect compatible with a balanced binomial distribution.  For
15 atoms the defect ratio is still about half, for 18 atoms it is only
a quarter of that expected from the uniform distribution. The ratio
then drops down to the percent range, indicating stable
self-organization, for 50 or more atoms.  Concerning the appearance of
defect atoms, Eq.\ (\ref{eq:N_thr}) gives $N>N_{\rm thr} \approx 40$.
Due to non-perfect bunching this threshold is shifted to somewhat
higher values.  The insert of Fig.\ \ref{fig:varn_defect} shows the
absolute number of defect atoms after 5 ms, averaged over the runs:
the minimum at $N\approx 60$ followed by a rise can be identified with
the threshold, which is in accordance with the previous, simple
estimate.

\begin{figure}[htbp]
  \centering
  \includegraphics[angle=270,width=7cm]{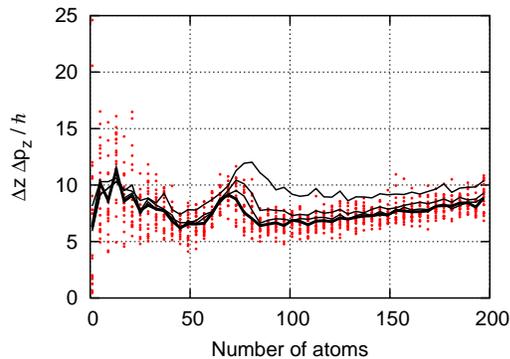}
  \caption{The normalized phase space volume of the system. Dots represent the 
    values taken at individual trajectories at 5 ms, the thick solid
    line is their average. The thin lines correspond to the average
    over the trajectories at 2, 3, and 4 ms. 
    The physical parameters are the same as in Fig.~\ref{fig:varn_defect}}
  \label{fig:varn_phdense}
\end{figure}

The transition from the perfectly ordered phase to the one where
defect atoms can be present manifests itself in the statistical
properties of the system.  As discussed in Sect.\ \ref{sec:selforg},
the appropriate measure of the thermal excitations is the phase space
volume, given by the Heisenberg uncertainty product $\Delta z \Delta
p_z / \hbar$.
In Fig.\ \ref{fig:varn_phdense}, this
is plotted as a function of the atom number: both the final phase
space volume of individual runs after 5 ms (dots) and the average over
these trajectories after 2, 3, 4 and 5 ms (continuous lines) are
shown.
For very few atoms, the phase space volume is scattered around the
value of $13.4$, corresponding to uniform spatial distribution and
mean kinetic energy $k_B T = \hbar \kappa$. For $N\approx 50$, 
the decrease of the average phase space volume and the
reducing variance around the mean show that the more stable the
self-organized pattern, the less thermally excited the system is.
At $N\approx 60$, a broad peak (due to the transition to a
double well potential) heralds the appearance of stable defect atoms.
This is followed by a slow but steady increase, which can be
attributed to the rising percentage of defects.

To indicate some of the dynamics, Fig.~\ref{fig:varn_phdense} shows
the Heisenberg product at earlier times as well: at 2, 3, and 4
milliseconds. The appearance of defect atoms slows down the
equilibration process, but the respective curves converge uniformly to
the one obtained at 5 ms. This demonstrates that the cooling time is
closely independent of the atom number.

\section{Collective effects in the self-organized phase}
\label{sec:collective}

\begin{figure}[htbp]
  \centering
  \includegraphics[angle=270,width=8cm]{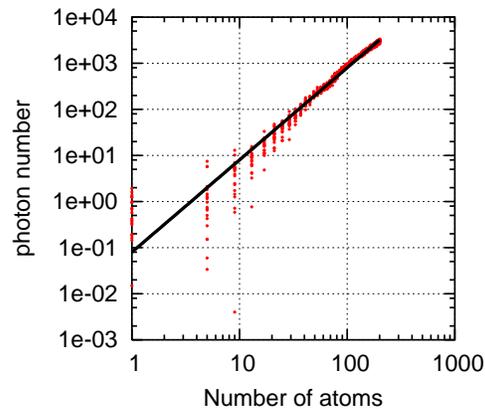}
  \caption{Photon number in the cavity 5ms after loading.  
    At high atom numbers ($N>20$) the simulation results (dots) are
    fitted well by a quadratic function (black line), indicating that
    the atoms scatter cooperatively.  The physical parameters are the
    same as in Fig.~\ref{fig:varn_defect}}
  \label{fig:suprad_a}
\end{figure}

The most direct evidence of cooperative action is superradiance into
the resonator mode that can be measured by detecting the power output
from the cavity.  In the self-organized checkerboard pattern each atom
radiates with the same phase, and so the cavity photon number
$|\alpha|^2$ increases quadratically with the number of atoms.  This
can be observed in the numerical simulations of Eq.(\ref{eq:sde}),
gradually increasing the number of atoms as detailed in the previous
Section.  The cavity photon number is shown on a logarithmic scale in
Fig.\ \ref{fig:suprad_a}.
For many atoms in the cavity ($N>20$) the intensity data are well
approximated by the fitted quadratic function $\abs{\alpha}^2=0.08 N^2$. 
The steady-state solution of Eq.\ (\ref{eq:sde_al})
reads
\begin{equation}
  \label{eq:superrad}
  |\alpha|^2 = I_0 \expect{\cos(kz_j)\cos(kx_j)}^2  N^2\;.
\end{equation}
If all atoms are perfectly at the antinodes 
($\expect{\cos(kz_j)\cos(kx_j)}^2 = \Theta = {\cal B} =1$)
this yields a coefficient of 0.125. The value from the simulations 
is $30$\% below this: the difference can be attributed to the
position spread and to the defect atoms.

\begin{figure}[htbp]
  \centering
  \includegraphics[angle=270,width=7cm]{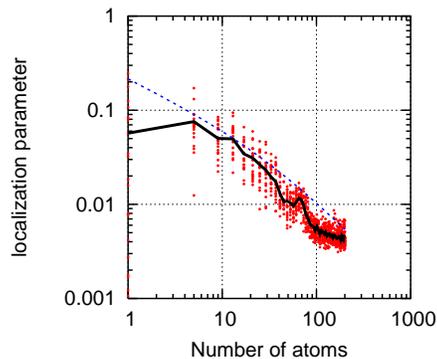}
  \caption{The localization parameter 
    ${\cal D}_z = \sum_i{\left(k z_i/\pi\right)^2}/N$ measured by the
    simulation (dots) and averaged over runs (continuous black line)
    shows an overall $1/N$ dependence as a function of the atom
    number. The dashed line is the approximation of Eq.\ 
    (\ref{eq:1perN}). The physical parameters are the same as in
    Fig.~\ref{fig:varn_defect}}
  \label{fig:varn_bunch}
\end{figure}

The superradiance has important effects on the spatial distribution of
the atoms about their respective field antinodes. With increasing
number of atoms the trap deepens so that the size of the atomic clouds
about the antinodes is compressed, as shown in
Fig.~\ref{fig:varn_bunch}.
We note again the appearance of the shoulder at $N\approx 60$ due to
the transition into the double-well potential with stable defect
atoms.  Apart from this, the overall decrease of the localization
parameter in the range $N=10, \ldots, 200$ is proportional to $1/N$. As we
show below, this scaling law can be derived by a careful examination
of the self-generated trap potential.

We consider one-dimensional motion of the atoms in the limit where
they are strongly localised in the vicinity of the antinodes.  In this
limit of harmonic oscillation the field amplitude in
Eq.~(\ref{eq:sde}a) is coupled to the atomic positions only through
the sum $\sum z_i^2$, i.e., through the localization parameter defined
in Eq.~(\ref{eq:localization}).  It is instructive to introduce new
coordinates in the configuration space: the mean radius $r
= \sqrt{\sum z_i^2/N}$, and a number of $N-1$ angular coordinates
$\varphi_j$ with canonically conjugate momenta $p_{\varphi j}=N m r^2
\dot\varphi_j$, these latter decoupled from the field dynamics.  Only
the radial motion is damped by the cavity cooling mechanism, the
angular ones are not.  For the coordinate $r$ the potential is
composed of the harmonic attraction $\propto r^2$ and a centrifugal
repulsion $\propto 1/r^2$.  There is a potential minimum and the
radius is damped into it by cavity cooling.  For large number of atoms
the cloud size at an antinode can be simply estimated by the radius at
the potential minimum. In this way we discard the role of fluctuations
in this coordinate, assuming that the variance is much smaller than
the mean.

For a quasi-stationary field amplitude, the harmonic potential is $m
\nu^2 N r^2/2$ with vibration frequency
 \begin{equation}
  \label{eq:harmonic}
  \nu = \sqrt{\frac{\hbar k^2}{m} \left( \frac{|U_0|\eta}{g} |\alpha| + 
2 |U_0| |\alpha|^2 \right)}\; .
\end{equation}
The an\-gu\-lar ki\-ne\-tic ener\-gies are of the form
$p_{\varphi_j}^2/(2 m N r^2)$, $j=1, \ldots, (N-1)$. The momenta
$p_{\varphi_j}$ can be estimated by their initial value in the
unorganized phase, when all degrees of freedom have the same energy
$k_B T/2$ and the radius is $r = \lambda/\sqrt{48}$. The potential
minimum is just at the radius where the harmonic potential energy
equals the sum of the centrifugal energies. Simple algebra leads to
the cloud size
\begin{equation}
  \label{eq:1perN}
  r^2 \approx \frac{\lambda^2}{8\pi \sqrt{3}} \sqrt{\frac{k_B T}{\hbar |U_0|}} 
\left( \frac{\eta}{g} |\alpha| + 
2 |\alpha|^2 \right)^{-\frac{1}{2}}\; ,
\end{equation}
and since $|\alpha|^2\propto N^2$, the $1/N$ law for the localization
parameter follows.  Altogether, the increase in atom number results in
a smaller cloud size in the vicinity of the antinodes, a very
important virtue of the collective atomic action. This compression is
limited by the dipole-dipole interaction of the atoms near the same
antinode, which effect was not taken into account in the present
model.

\section{Conclusion}
\label{sect:conclusion}

A dilute cloud of non-interacting cold atoms in a high-Q cavity can
undergo a phase transition if driven from the side by a laser
sufficiently red detuned from an atomic resonance.  Increasing the
laser power above a threshold the atoms self-organize into a lattice
so as to scatter most effectively into the cavity mode. In this way
the atoms minimize their energy in the optical potential generated
from the interference pattern of the cavity and pump field. Under
proper conditions the atoms are cooled in this process giving long
term stability to the pattern.

The phenomenon has been previously seen in simulations and strong
evidence was found in experiments \cite{black03}, but threshold,
scaling and efficiency of the effect with atom number, cavity
parameters and system size remained unclear.  Here we gave a thorough
analytical discussion of the effect.  A continuous density approach
allowed us to derive analytical formulas for the critical pump power
as a function of atom number and cavity volume and showed that the
effect should persist if one scales up the volume at fixed atomic
density, i.e., $Ng^2=\textrm{const}$. Numerical simulations of the
evolution for finite durations revealed a hysteresis between the
ordered and homogeneous density phase on varying the control
parameter, i.e., the pumping strength.  This shifts the observable
threshold in the pumping strength, which then scales with $Ng^4$. 
We still have a cooling mechanism for large numbers of any type of
optically polarizable particles. 

The system is composed of non-interacting atoms that communicate via a
commonly coupled, single cavity mode. The cavity component of the
system plays a multiple role.  First, the binding energy of the
ordered phase is stored in the superradiantly enhanced field intensity
of this single mode. Next, as an attractive feature of this system,
the outcoupled field intensity directly serves as a possibility of
time-resolved monitoring of the formation of the ordered phase. Note
that in setups without resonator, the uncontrolled scattered field can
lead to a binding of micrometer-sized particles in an ordered pattern
in liquid \cite{burns89,singer03}. In the cavity scheme, finally, the
viscous motion is provided by the dynamically coupled single-mode
cavity field. Apart from the geometry, this is also a distinctive
feature with respect to CARL.

From the laser cooling point of view, the picture is complicated by
the phase transition. The appropriate measure of the cooling
efficiency is the phase space volume occupied by the system. The
self-organization process reduces this volume below that of a system
at the cavity cooling limit with uniform spatial distribution.
Moreover, the steady state is established within milliseconds, and
this is closely independent of the number of atoms in the cloud
(numerical simulations confirm this up to few hundreds of atoms).
Collective cooling was previously known only for the stochastic
cooling method \cite{raizen98} and for common vibrational modes of
trapped particles \cite{beige05}.  The collective behaviour strongly
improves the localization, i.e., the size of the atom cloud pieces at
the trapping sites is proportional to the inverse of the atom number.

This work could be extended to various directions. First of all, the
numerical simulations should confirm some of the statements of the
present paper on a larger range of the atom number, or on a longer
time scale (e.g., the dependence of the hysteresis on the duration of
the evolution). Our prescription for the pump-cavity detuning
$\Delta_C$ in Eq.\ (\ref{eq:dc}) is probably impractical in the limit
of large atom numbers, as one has to go closer to the resonance in
order to initiate the self-organization.  Next, the maximum achiveable
densities can not be determined in the present model as some of the
limiting factors were omitted, e.g., vacum-mediated dipole-dipole
interaction between the atoms, the eventual superradiant enhancement
of the spontaneous scattering into lateral directions, etc. In
extremely good cavities, quantum effects in the motion of the atoms
begin to play an important role \cite{zippilli05,vukics05}, which was
not studied here.  Finally, the nature of the phase transition is
strongly determined by the geometry of the cavity mode: the possible
trapping sites are defined by the antinodes of the cavity mode and the
transverse pumping field. This situation can be essentially modified
in a cavities with different geometry, e.g., ring or confocal
resonator.

We thank Walter Rantner for useful discussions and Zolt\'an Kurucz for
reading the manuscript. This work was supported by the National
Scientific Fund of Hungary (Contract Nos.~T043079, T049234), the
Bolyai Program of the Hungarian Academy of Sciences, and by the
Austrian FWF SFB P12.


\begin{thebibliography}{45}
\expandafter\ifx\csname natexlab\endcsname\relax\def\natexlab#1{#1}\fi
\expandafter\ifx\csname bibnamefont\endcsname\relax
  \def\bibnamefont#1{#1}\fi
\expandafter\ifx\csname bibfnamefont\endcsname\relax
  \def\bibfnamefont#1{#1}\fi
\expandafter\ifx\csname citenamefont\endcsname\relax
  \def\citenamefont#1{#1}\fi
\expandafter\ifx\csname url\endcsname\relax
  \def\url#1{\texttt{#1}}\fi
\expandafter\ifx\csname urlprefix\endcsname\relax\def\urlprefix{URL }\fi
\providecommand{\bibinfo}[2]{#2}
\providecommand{\eprint}[2][]{\url{#2}}

\bibitem[{\citenamefont{Metcalf and van~der Straten}(1999)}]{metcalf}
\bibinfo{author}{\bibfnamefont{H.~J.} \bibnamefont{Metcalf}} \bibnamefont{and}
  \bibinfo{author}{\bibfnamefont{P.}~\bibnamefont{van~der Straten}},
  \emph{\bibinfo{title}{Laser Cooling and Trapping}}
  (\bibinfo{publisher}{Springer, New York}, \bibinfo{year}{1999}).

\bibitem[{\citenamefont{Meystre}(2001)}]{meystre}
\bibinfo{author}{\bibfnamefont{P.}~\bibnamefont{Meystre}},
  \emph{\bibinfo{title}{Atom Optics}} (\bibinfo{publisher}{Springer, New York},
  \bibinfo{year}{2001}).

\bibitem[{\citenamefont{Castin}(2001)}]{castin}
\bibinfo{author}{\bibfnamefont{Y.}~\bibnamefont{Castin}}, in
  \emph{\bibinfo{booktitle}{Coherent atomic matter waves, Lecture Notes of Les
  Houches Summer School}}, edited by
  \bibinfo{editor}{\bibfnamefont{R.}~\bibnamefont{Kaiser}},
  \bibinfo{editor}{\bibfnamefont{C.}~\bibnamefont{Westbrook}},
  \bibnamefont{and} \bibinfo{editor}{\bibfnamefont{F.}~\bibnamefont{David}}
  (\bibinfo{publisher}{EDP Sciences and Springer-Verlag},
  \bibinfo{year}{2001}), pp. \bibinfo{pages}{1--136}.

\bibitem[{\citenamefont{Morice et~al.}(1995)\citenamefont{Morice, Castin, and
  Dalibard}}]{morice95}
\bibinfo{author}{\bibfnamefont{O.}~\bibnamefont{Morice}},
  \bibinfo{author}{\bibfnamefont{Y.}~\bibnamefont{Castin}}, \bibnamefont{and}
  \bibinfo{author}{\bibfnamefont{J.}~\bibnamefont{Dalibard}},
  \bibinfo{journal}{Phys. Rev. A} \textbf{\bibinfo{volume}{51}},
  \bibinfo{pages}{3896} (\bibinfo{year}{1995}).

\bibitem[{\citenamefont{Lagendijk et~al.}(1997)\citenamefont{Lagendijk,
  Nienhuis, van Tiggelen, and de~Vries}}]{lagendijk97}
\bibinfo{author}{\bibfnamefont{A.}~\bibnamefont{Lagendijk}},
  \bibinfo{author}{\bibfnamefont{B.}~\bibnamefont{Nienhuis}},
  \bibinfo{author}{\bibfnamefont{B.~A.} \bibnamefont{van Tiggelen}},
  \bibnamefont{and} \bibinfo{author}{\bibfnamefont{P.}~\bibnamefont{de~Vries}},
  \bibinfo{journal}{Phys. Rev. Lett.} \textbf{\bibinfo{volume}{79}},
  \bibinfo{pages}{657} (\bibinfo{year}{1997}).

\bibitem[{\citenamefont{Labeyrie
  et~al.}(2003{\natexlab{a}})\citenamefont{Labeyrie, Vaujour, M\"uller,
  Delande, Miniatura, Wilkowski, and Kaiser}}]{labeyrie03a}
\bibinfo{author}{\bibfnamefont{G.}~\bibnamefont{Labeyrie}},
  \bibinfo{author}{\bibfnamefont{E.}~\bibnamefont{Vaujour}},
  \bibinfo{author}{\bibfnamefont{C.~A.} \bibnamefont{M\"uller}},
  \bibinfo{author}{\bibfnamefont{D.}~\bibnamefont{Delande}},
  \bibinfo{author}{\bibfnamefont{C.}~\bibnamefont{Miniatura}},
  \bibinfo{author}{\bibfnamefont{D.}~\bibnamefont{Wilkowski}},
  \bibnamefont{and} \bibinfo{author}{\bibfnamefont{R.}~\bibnamefont{Kaiser}},
  \bibinfo{journal}{Phys. Rev. Lett.} \textbf{\bibinfo{volume}{91}},
  \bibinfo{pages}{223904} (\bibinfo{year}{2003}{\natexlab{a}}).

\bibitem[{\citenamefont{Labeyrie
  et~al.}(2003{\natexlab{b}})\citenamefont{Labeyrie, Delande, M\"uller,
  Miniatura, and Kaiser}}]{labeyrie03b}
\bibinfo{author}{\bibfnamefont{G.}~\bibnamefont{Labeyrie}},
  \bibinfo{author}{\bibfnamefont{D.}~\bibnamefont{Delande}},
  \bibinfo{author}{\bibfnamefont{C.~A.} \bibnamefont{M\"uller}},
  \bibinfo{author}{\bibfnamefont{C.}~\bibnamefont{Miniatura}},
  \bibnamefont{and} \bibinfo{author}{\bibfnamefont{R.}~\bibnamefont{Kaiser}},
  \bibinfo{journal}{Phys. Rev. A} \textbf{\bibinfo{volume}{67}},
  \bibinfo{pages}{033814} (\bibinfo{year}{2003}{\natexlab{b}}).

\bibitem[{\citenamefont{Sesko et~al.}(1991)\citenamefont{Sesko, Walker, and
  Wieman}}]{sesko91}
\bibinfo{author}{\bibfnamefont{D.~W.} \bibnamefont{Sesko}},
  \bibinfo{author}{\bibfnamefont{T.~G.} \bibnamefont{Walker}},
  \bibnamefont{and} \bibinfo{author}{\bibfnamefont{C.~E.}
  \bibnamefont{Wieman}}, \bibinfo{journal}{J. Opt. Soc. Am. B}
  \textbf{\bibinfo{volume}{8}}, \bibinfo{pages}{946} (\bibinfo{year}{1991}).

\bibitem[{\citenamefont{Hood et~al.}(2000)\citenamefont{Hood, Lynn, Doherty,
  and Kimble}}]{hood00}
\bibinfo{author}{\bibfnamefont{C.~J.} \bibnamefont{Hood}},
  \bibinfo{author}{\bibfnamefont{T.~W.} \bibnamefont{Lynn}},
  \bibinfo{author}{\bibfnamefont{A.~C.} \bibnamefont{Doherty}},
  \bibnamefont{and} \bibinfo{author}{\bibfnamefont{A.~S. P. H.~J.}
  \bibnamefont{Kimble}}, \bibinfo{journal}{Science}
  \textbf{\bibinfo{volume}{287}}, \bibinfo{pages}{1447} (\bibinfo{year}{2000}).

\bibitem[{\citenamefont{Pinkse et~al.}(2000)\citenamefont{Pinkse, Fischer,
  Maunz, and Rempe}}]{pinkse00}
\bibinfo{author}{\bibfnamefont{P.~W.~H.} \bibnamefont{Pinkse}},
  \bibinfo{author}{\bibfnamefont{T.}~\bibnamefont{Fischer}},
  \bibinfo{author}{\bibfnamefont{P.}~\bibnamefont{Maunz}}, \bibnamefont{and}
  \bibinfo{author}{\bibfnamefont{G.}~\bibnamefont{Rempe}},
  \bibinfo{journal}{Nature} \textbf{\bibinfo{volume}{404}},
  \bibinfo{pages}{365} (\bibinfo{year}{2000}).

\bibitem[{\citenamefont{Horak et~al.}(1997)\citenamefont{Horak, Hechenblaikner,
  Gheri, Stecher, and Ritsch}}]{horak97}
\bibinfo{author}{\bibfnamefont{P.}~\bibnamefont{Horak}},
  \bibinfo{author}{\bibfnamefont{G.}~\bibnamefont{Hechenblaikner}},
  \bibinfo{author}{\bibfnamefont{K.~M.} \bibnamefont{Gheri}},
  \bibinfo{author}{\bibfnamefont{H.}~\bibnamefont{Stecher}}, \bibnamefont{and}
  \bibinfo{author}{\bibfnamefont{H.}~\bibnamefont{Ritsch}},
  \bibinfo{journal}{Phys. Rev. Lett.} \textbf{\bibinfo{volume}{79}},
  \bibinfo{pages}{4974} (\bibinfo{year}{1997}).

\bibitem[{\citenamefont{Domokos and Ritsch}(2003)}]{domokos03}
\bibinfo{author}{\bibfnamefont{P.}~\bibnamefont{Domokos}} \bibnamefont{and}
  \bibinfo{author}{\bibfnamefont{H.}~\bibnamefont{Ritsch}},
  \bibinfo{journal}{J. Opt. Soc. Am. B} \textbf{\bibinfo{volume}{20}},
  \bibinfo{pages}{1098} (\bibinfo{year}{2003}).

\bibitem[{\citenamefont{Maunz et~al.}(2004)\citenamefont{Maunz, Puppe,
  Schuster, Syassen, Pinkse, and Rempe}}]{maunz04}
\bibinfo{author}{\bibfnamefont{P.}~\bibnamefont{Maunz}},
  \bibinfo{author}{\bibfnamefont{T.}~\bibnamefont{Puppe}},
  \bibinfo{author}{\bibfnamefont{I.}~\bibnamefont{Schuster}},
  \bibinfo{author}{\bibfnamefont{N.}~\bibnamefont{Syassen}},
  \bibinfo{author}{\bibfnamefont{P.~W.~H.} \bibnamefont{Pinkse}},
  \bibnamefont{and} \bibinfo{author}{\bibfnamefont{G.}~\bibnamefont{Rempe}},
  \bibinfo{journal}{Nature} \textbf{\bibinfo{volume}{428}}, \bibinfo{pages}{50}
  (\bibinfo{year}{2004}).

\bibitem[{\citenamefont{Nussmann et~al.}(2005)\citenamefont{Nussmann, Murr,
  Hijlkema, Weber, Kuhn, and Rempe}}]{nussmann05}
\bibinfo{author}{\bibfnamefont{S.}~\bibnamefont{Nussmann}},
  \bibinfo{author}{\bibfnamefont{K.}~\bibnamefont{Murr}},
  \bibinfo{author}{\bibfnamefont{M.}~\bibnamefont{Hijlkema}},
  \bibinfo{author}{\bibfnamefont{B.}~\bibnamefont{Weber}},
  \bibinfo{author}{\bibfnamefont{A.}~\bibnamefont{Kuhn}}, \bibnamefont{and}
  \bibinfo{author}{\bibfnamefont{G.}~\bibnamefont{Rempe}},
  \bibinfo{journal}{quant-ph/0506067}  (\bibinfo{year}{2005}).

\bibitem[{\citenamefont{M\"unstermann et~al.}(2000)\citenamefont{M\"unstermann,
  Fischer, Maunz, Pinkse, and Rempe}}]{munstermann00}
\bibinfo{author}{\bibfnamefont{P.}~\bibnamefont{M\"unstermann}},
  \bibinfo{author}{\bibfnamefont{T.}~\bibnamefont{Fischer}},
  \bibinfo{author}{\bibfnamefont{P.}~\bibnamefont{Maunz}},
  \bibinfo{author}{\bibfnamefont{P.~W.~H.} \bibnamefont{Pinkse}},
  \bibnamefont{and} \bibinfo{author}{\bibfnamefont{G.}~\bibnamefont{Rempe}},
  \bibinfo{journal}{Phys. Rev. Lett.} \textbf{\bibinfo{volume}{84}},
  \bibinfo{pages}{4068} (\bibinfo{year}{2000}).

\bibitem[{\citenamefont{Black et~al.}(2005)\citenamefont{Black, Thompson, and
  Vuleti\'c}}]{black05}
\bibinfo{author}{\bibfnamefont{A.~T.} \bibnamefont{Black}},
  \bibinfo{author}{\bibfnamefont{J.~K.} \bibnamefont{Thompson}},
  \bibnamefont{and}
  \bibinfo{author}{\bibfnamefont{V.}~\bibnamefont{Vuleti\'c}},
  \bibinfo{journal}{Journal of Physics B: At. Mol. Opt. Phys.}
  \textbf{\bibinfo{volume}{38}}, \bibinfo{pages}{S605} (\bibinfo{year}{2005}).

\bibitem[{\citenamefont{Asboth et~al.}(2004)\citenamefont{Asboth, Domokos, and
  Ritsch}}]{asboth04}
\bibinfo{author}{\bibfnamefont{J.~K.} \bibnamefont{Asboth}},
  \bibinfo{author}{\bibfnamefont{P.}~\bibnamefont{Domokos}}, \bibnamefont{and}
  \bibinfo{author}{\bibfnamefont{H.}~\bibnamefont{Ritsch}},
  \bibinfo{journal}{Phys. Rev. A} \textbf{\bibinfo{volume}{70}},
  \bibinfo{pages}{013414} (\bibinfo{year}{2004}).

\bibitem[{\citenamefont{Fischer et~al.}(2001)\citenamefont{Fischer, Maunz,
  Puppe, Pinkse, and Rempe}}]{fischer01}
\bibinfo{author}{\bibfnamefont{T.}~\bibnamefont{Fischer}},
  \bibinfo{author}{\bibfnamefont{P.}~\bibnamefont{Maunz}},
  \bibinfo{author}{\bibfnamefont{T.}~\bibnamefont{Puppe}},
  \bibinfo{author}{\bibfnamefont{P.~W.~H.} \bibnamefont{Pinkse}},
  \bibnamefont{and} \bibinfo{author}{\bibfnamefont{G.}~\bibnamefont{Rempe}},
  \bibinfo{journal}{New J. of Phys.} \textbf{\bibinfo{volume}{3}},
  \bibinfo{pages}{11.1} (\bibinfo{year}{2001}).

\bibitem[{\citenamefont{Horak and Ritsch}(2001)}]{horak01}
\bibinfo{author}{\bibfnamefont{P.}~\bibnamefont{Horak}} \bibnamefont{and}
  \bibinfo{author}{\bibfnamefont{H.}~\bibnamefont{Ritsch}},
  \bibinfo{journal}{Phys. Rev. A} \textbf{\bibinfo{volume}{64}},
  \bibinfo{pages}{033422} (\bibinfo{year}{2001}).

\bibitem[{\citenamefont{Domokos and Ritsch}(2002)}]{domokos02b}
\bibinfo{author}{\bibfnamefont{P.}~\bibnamefont{Domokos}} \bibnamefont{and}
  \bibinfo{author}{\bibfnamefont{H.}~\bibnamefont{Ritsch}},
  \bibinfo{journal}{Phys. Rev. Lett.} \textbf{\bibinfo{volume}{89}},
  \bibinfo{pages}{253003} (\bibinfo{year}{2002}).

\bibitem[{\citenamefont{Black et~al.}(2003)\citenamefont{Black, Chan, and
  Vuleti\'c}}]{black03}
\bibinfo{author}{\bibfnamefont{A.~T.} \bibnamefont{Black}},
  \bibinfo{author}{\bibfnamefont{H.~W.} \bibnamefont{Chan}}, \bibnamefont{and}
  \bibinfo{author}{\bibfnamefont{V.}~\bibnamefont{Vuleti\'c}},
  \bibinfo{journal}{Phys. Rev. Lett.} \textbf{\bibinfo{volume}{91}},
  \bibinfo{pages}{203001} (\bibinfo{year}{2003}).

\bibitem[{\citenamefont{Kruse et~al.}(2003)\citenamefont{Kruse, von Cube,
  Zimmermann, and Courteille}}]{kruse03}
\bibinfo{author}{\bibfnamefont{D.}~\bibnamefont{Kruse}},
  \bibinfo{author}{\bibfnamefont{C.}~\bibnamefont{von Cube}},
  \bibinfo{author}{\bibfnamefont{C.}~\bibnamefont{Zimmermann}},
  \bibnamefont{and} \bibinfo{author}{\bibfnamefont{P.~W.}
  \bibnamefont{Courteille}}, \bibinfo{journal}{Phys. Rev. Lett.}
  \textbf{\bibinfo{volume}{91}}, \bibinfo{pages}{183601}
  (\bibinfo{year}{2003}).

\bibitem[{\citenamefont{Nagorny et~al.}(2003)\citenamefont{Nagorny, Els\"asser,
  and Hemmerich}}]{nagorny03}
\bibinfo{author}{\bibfnamefont{B.}~\bibnamefont{Nagorny}},
  \bibinfo{author}{\bibfnamefont{T.}~\bibnamefont{Els\"asser}},
  \bibnamefont{and}
  \bibinfo{author}{\bibfnamefont{A.}~\bibnamefont{Hemmerich}},
  \bibinfo{journal}{Phys. Rev. Lett.} \textbf{\bibinfo{volume}{91}},
  \bibinfo{pages}{153003} (\bibinfo{year}{2003}).

\bibitem[{\citenamefont{Elsässer et~al.}(2004)\citenamefont{Elsässer, Nagorny,
  and Hemmerich}}]{elsasser04}
\bibinfo{author}{\bibfnamefont{T.}~\bibnamefont{Elsässer}},
  \bibinfo{author}{\bibfnamefont{B.}~\bibnamefont{Nagorny}}, \bibnamefont{and}
  \bibinfo{author}{\bibfnamefont{A.}~\bibnamefont{Hemmerich}},
  \bibinfo{journal}{Phys. Rev. A} \textbf{\bibinfo{volume}{69}},
  \bibinfo{pages}{033403} (\bibinfo{year}{2004}).

\bibitem[{\citenamefont{Slama et~al.}(2005)\citenamefont{Slama, von Cube, Deh,
  Ludewig, Zimmermann, and Courteille}}]{slama04}
\bibinfo{author}{\bibfnamefont{S.}~\bibnamefont{Slama}},
  \bibinfo{author}{\bibfnamefont{C.}~\bibnamefont{von Cube}},
  \bibinfo{author}{\bibfnamefont{B.}~\bibnamefont{Deh}},
  \bibinfo{author}{\bibfnamefont{A.}~\bibnamefont{Ludewig}},
  \bibinfo{author}{\bibfnamefont{C.}~\bibnamefont{Zimmermann}},
  \bibnamefont{and} \bibinfo{author}{\bibfnamefont{P.~W.}
  \bibnamefont{Courteille}}, \bibinfo{journal}{Phys. Rev. Lett.}
  \textbf{\bibinfo{volume}{94}}, \bibinfo{pages}{193901}
  (\bibinfo{year}{2005}).

\bibitem[{\citenamefont{Bonifacio et~al.}(1994)\citenamefont{Bonifacio,
  DeSalvo, Narducci, and D'Angelo}}]{bonifacio94}
\bibinfo{author}{\bibfnamefont{R.}~\bibnamefont{Bonifacio}},
  \bibinfo{author}{\bibfnamefont{L.}~\bibnamefont{DeSalvo}},
  \bibinfo{author}{\bibfnamefont{L.~M.} \bibnamefont{Narducci}},
  \bibnamefont{and} \bibinfo{author}{\bibfnamefont{E.~J.}
  \bibnamefont{D'Angelo}}, \bibinfo{journal}{Phys. Rev. A}
  \textbf{\bibinfo{volume}{50}}, \bibinfo{pages}{1716} (\bibinfo{year}{1994}).

\bibitem[{\citenamefont{Javaloyes et~al.}(2004)\citenamefont{Javaloyes, Perrin,
  Lippi, and Politi}}]{javaloyes04}
\bibinfo{author}{\bibfnamefont{J.}~\bibnamefont{Javaloyes}},
  \bibinfo{author}{\bibfnamefont{M.}~\bibnamefont{Perrin}},
  \bibinfo{author}{\bibfnamefont{G.~L.} \bibnamefont{Lippi}}, \bibnamefont{and}
  \bibinfo{author}{\bibfnamefont{A.}~\bibnamefont{Politi}},
  \bibinfo{journal}{Phys. Rev. A} \textbf{\bibinfo{volume}{70}},
  \bibinfo{pages}{023405} (\bibinfo{year}{2004}).

\bibitem[{\citenamefont{von Cube et~al.}(2004)\citenamefont{von Cube, Slama,
  Kruse, Zimmermann, Courteille, Robb, Piovella, and Bonifacio}}]{cube04}
\bibinfo{author}{\bibfnamefont{C.}~\bibnamefont{von Cube}},
  \bibinfo{author}{\bibfnamefont{S.}~\bibnamefont{Slama}},
  \bibinfo{author}{\bibfnamefont{D.}~\bibnamefont{Kruse}},
  \bibinfo{author}{\bibfnamefont{C.}~\bibnamefont{Zimmermann}},
  \bibinfo{author}{\bibfnamefont{P.~W.} \bibnamefont{Courteille}},
  \bibinfo{author}{\bibfnamefont{G.~R.~M.} \bibnamefont{Robb}},
  \bibinfo{author}{\bibfnamefont{N.}~\bibnamefont{Piovella}}, \bibnamefont{and}
  \bibinfo{author}{\bibfnamefont{R.}~\bibnamefont{Bonifacio}},
  \bibinfo{journal}{Phys. Rev. Lett.} \textbf{\bibinfo{volume}{93}},
  \bibinfo{pages}{083601} (\bibinfo{year}{2004}).

\bibitem[{\citenamefont{Robb et~al.}(2004)\citenamefont{Robb, Piovella,
  Ferraro, Bonifacio, Courteille, and Zimmermann}}]{robb04}
\bibinfo{author}{\bibfnamefont{G.~R.~M.} \bibnamefont{Robb}},
  \bibinfo{author}{\bibfnamefont{N.}~\bibnamefont{Piovella}},
  \bibinfo{author}{\bibfnamefont{A.}~\bibnamefont{Ferraro}},
  \bibinfo{author}{\bibfnamefont{R.}~\bibnamefont{Bonifacio}},
  \bibinfo{author}{\bibfnamefont{P.~W.} \bibnamefont{Courteille}},
  \bibnamefont{and}
  \bibinfo{author}{\bibfnamefont{C.}~\bibnamefont{Zimmermann}},
  \bibinfo{journal}{Phys. Rev. A} \textbf{\bibinfo{volume}{69}},
  \bibinfo{pages}{041403} (\bibinfo{year}{2004}).

\bibitem[{\citenamefont{Sauer et~al.}(2004)\citenamefont{Sauer, Fortier, Chang,
  Hamley, and Chapman}}]{sauer04}
\bibinfo{author}{\bibfnamefont{J.~A.} \bibnamefont{Sauer}},
  \bibinfo{author}{\bibfnamefont{K.~M.} \bibnamefont{Fortier}},
  \bibinfo{author}{\bibfnamefont{M.~S.} \bibnamefont{Chang}},
  \bibinfo{author}{\bibfnamefont{C.~D.} \bibnamefont{Hamley}},
  \bibnamefont{and} \bibinfo{author}{\bibfnamefont{M.~S.}
  \bibnamefont{Chapman}}, \bibinfo{journal}{Phys. Rev. A}
  \textbf{\bibinfo{volume}{69}}, \bibinfo{pages}{051804(R)}
  (\bibinfo{year}{2004}).

\bibitem[{\citenamefont{Schrader et~al.}(2001)\citenamefont{Schrader, Kuhr,
  Alt, M\"uller, Gomer, and Meschede}}]{schrader01}
\bibinfo{author}{\bibfnamefont{D.}~\bibnamefont{Schrader}},
  \bibinfo{author}{\bibfnamefont{S.}~\bibnamefont{Kuhr}},
  \bibinfo{author}{\bibfnamefont{W.}~\bibnamefont{Alt}},
  \bibinfo{author}{\bibfnamefont{M.}~\bibnamefont{M\"uller}},
  \bibinfo{author}{\bibfnamefont{V.}~\bibnamefont{Gomer}}, \bibnamefont{and}
  \bibinfo{author}{\bibfnamefont{D.}~\bibnamefont{Meschede}},
  \bibinfo{journal}{Appl. Phys. B} \textbf{\bibinfo{volume}{73}},
  \bibinfo{pages}{819} (\bibinfo{year}{2001}).

\bibitem[{\citenamefont{Schrader et~al.}(2004)\citenamefont{Schrader, Dotsenko,
  Khudaverdyan, Miroshnychenko, Rauschenbeutel, and Meschede}}]{schrader04}
\bibinfo{author}{\bibfnamefont{D.}~\bibnamefont{Schrader}},
  \bibinfo{author}{\bibfnamefont{I.}~\bibnamefont{Dotsenko}},
  \bibinfo{author}{\bibfnamefont{M.}~\bibnamefont{Khudaverdyan}},
  \bibinfo{author}{\bibfnamefont{Y.}~\bibnamefont{Miroshnychenko}},
  \bibinfo{author}{\bibfnamefont{A.}~\bibnamefont{Rauschenbeutel}},
  \bibnamefont{and} \bibinfo{author}{\bibfnamefont{D.}~\bibnamefont{Meschede}},
  \bibinfo{journal}{Phys. Rev. Lett.} \textbf{\bibinfo{volume}{93}},
  \bibinfo{pages}{150501} (\bibinfo{year}{2004}).

\bibitem[{\citenamefont{Domokos et~al.}(2001)\citenamefont{Domokos, Horak, and
  Ritsch}}]{domokos01}
\bibinfo{author}{\bibfnamefont{P.}~\bibnamefont{Domokos}},
  \bibinfo{author}{\bibfnamefont{P.}~\bibnamefont{Horak}}, \bibnamefont{and}
  \bibinfo{author}{\bibfnamefont{H.}~\bibnamefont{Ritsch}},
  \bibinfo{journal}{Journal of Physics B: At. Mol. Opt. Phys.}
  \textbf{\bibinfo{volume}{34}}, \bibinfo{pages}{187} (\bibinfo{year}{2001}).

\bibitem[{\citenamefont{Domokos et~al.}(2004)\citenamefont{Domokos, Vukics, and
  Ritsch}}]{domokos04}
\bibinfo{author}{\bibfnamefont{P.}~\bibnamefont{Domokos}},
  \bibinfo{author}{\bibfnamefont{A.}~\bibnamefont{Vukics}}, \bibnamefont{and}
  \bibinfo{author}{\bibfnamefont{H.}~\bibnamefont{Ritsch}},
  \bibinfo{journal}{Phys. Rev. Lett.} \textbf{\bibinfo{volume}{92}},
  \bibinfo{pages}{103601} (\bibinfo{year}{2004}).

\bibitem[{\citenamefont{Hechenblaikner
  et~al.}(1998)\citenamefont{Hechenblaikner, Gangl, Horak, and
  Ritsch}}]{hechenblaikner98}
\bibinfo{author}{\bibfnamefont{G.}~\bibnamefont{Hechenblaikner}},
  \bibinfo{author}{\bibfnamefont{M.}~\bibnamefont{Gangl}},
  \bibinfo{author}{\bibfnamefont{P.}~\bibnamefont{Horak}}, \bibnamefont{and}
  \bibinfo{author}{\bibfnamefont{H.}~\bibnamefont{Ritsch}},
  \bibinfo{journal}{Phys. Rev. A} \textbf{\bibinfo{volume}{58}},
  \bibinfo{pages}{3030} (\bibinfo{year}{1998}).

\bibitem[{\citenamefont{Doherty et~al.}(2000)\citenamefont{Doherty, Lynn, Hood,
  and Kimble}}]{doherty00}
\bibinfo{author}{\bibfnamefont{A.~C.} \bibnamefont{Doherty}},
  \bibinfo{author}{\bibfnamefont{T.~W.} \bibnamefont{Lynn}},
  \bibinfo{author}{\bibfnamefont{C.~J.} \bibnamefont{Hood}}, \bibnamefont{and}
  \bibinfo{author}{\bibfnamefont{H.~J.} \bibnamefont{Kimble}},
  \bibinfo{journal}{Phys. Rev. A} \textbf{\bibinfo{volume}{63}},
  \bibinfo{pages}{013401} (\bibinfo{year}{2000}).

\bibitem[{\citenamefont{Vuleti\'c et~al.}(2001)\citenamefont{Vuleti\'c, Chan,
  and Black}}]{vuletic01}
\bibinfo{author}{\bibfnamefont{V.}~\bibnamefont{Vuleti\'c}},
  \bibinfo{author}{\bibfnamefont{H.~W.} \bibnamefont{Chan}}, \bibnamefont{and}
  \bibinfo{author}{\bibfnamefont{A.~T.} \bibnamefont{Black}},
  \bibinfo{journal}{Phys. Rev. A} \textbf{\bibinfo{volume}{64}},
  \bibinfo{pages}{033405} (\bibinfo{year}{2001}).

\bibitem[{\citenamefont{van Enk et~al.}(2001)\citenamefont{van Enk, McKeever,
  Kimble, and Ye}}]{vanenk01}
\bibinfo{author}{\bibfnamefont{S.~J.} \bibnamefont{van Enk}},
  \bibinfo{author}{\bibfnamefont{J.}~\bibnamefont{McKeever}},
  \bibinfo{author}{\bibfnamefont{H.~J.} \bibnamefont{Kimble}},
  \bibnamefont{and} \bibinfo{author}{\bibfnamefont{J.}~\bibnamefont{Ye}},
  \bibinfo{journal}{Phys. Rev. A} \textbf{\bibinfo{volume}{64}},
  \bibinfo{pages}{013407} (\bibinfo{year}{2001}).

\bibitem[{\citenamefont{Murr}(2003)}]{murr03}
\bibinfo{author}{\bibfnamefont{K.}~\bibnamefont{Murr}},
  \bibinfo{journal}{Journal of Physics B: At. Mol. Opt. Phys.}
  \textbf{\bibinfo{volume}{36}}, \bibinfo{pages}{2515} (\bibinfo{year}{2003}).

\bibitem[{\citenamefont{Burns et~al.}(1989)\citenamefont{Burns, Fournier, and
  Golovchenko}}]{burns89}
\bibinfo{author}{\bibfnamefont{M.~M.} \bibnamefont{Burns}},
  \bibinfo{author}{\bibfnamefont{J.}~\bibnamefont{Fournier}}, \bibnamefont{and}
  \bibinfo{author}{\bibfnamefont{J.~A.} \bibnamefont{Golovchenko}},
  \bibinfo{journal}{PRL}  (\bibinfo{year}{1989}).

\bibitem[{\citenamefont{Singer et~al.}(2003)\citenamefont{Singer, Frick,
  Bernet, and Ritsch-Marte}}]{singer03}
\bibinfo{author}{\bibfnamefont{W.}~\bibnamefont{Singer}},
  \bibinfo{author}{\bibfnamefont{M.}~\bibnamefont{Frick}},
  \bibinfo{author}{\bibfnamefont{S.}~\bibnamefont{Bernet}}, \bibnamefont{and}
  \bibinfo{author}{\bibfnamefont{M.}~\bibnamefont{Ritsch-Marte}},
  \bibinfo{journal}{J. Opt. Soc. Am. B} \textbf{\bibinfo{volume}{20}},
  \bibinfo{pages}{1568} (\bibinfo{year}{2003}).

\bibitem[{\citenamefont{Raizen et~al.}(1998)\citenamefont{Raizen, Koga,
  Sundaram, Kishimoto, Takuma, and Tajima}}]{raizen98}
\bibinfo{author}{\bibfnamefont{M.~G.} \bibnamefont{Raizen}},
  \bibinfo{author}{\bibfnamefont{J.}~\bibnamefont{Koga}},
  \bibinfo{author}{\bibfnamefont{B.}~\bibnamefont{Sundaram}},
  \bibinfo{author}{\bibfnamefont{Y.}~\bibnamefont{Kishimoto}},
  \bibinfo{author}{\bibfnamefont{H.}~\bibnamefont{Takuma}}, \bibnamefont{and}
  \bibinfo{author}{\bibfnamefont{T.}~\bibnamefont{Tajima}},
  \bibinfo{journal}{Phys. Rev. A} \textbf{\bibinfo{volume}{58}},
  \bibinfo{pages}{4757} (\bibinfo{year}{1998}).

\bibitem[{\citenamefont{Beige et~al.}(2005)\citenamefont{Beige, Knight, and
  Vitiello}}]{beige05}
\bibinfo{author}{\bibfnamefont{A.}~\bibnamefont{Beige}},
  \bibinfo{author}{\bibfnamefont{P.~L.} \bibnamefont{Knight}},
  \bibnamefont{and} \bibinfo{author}{\bibfnamefont{G.}~\bibnamefont{Vitiello}},
  \bibinfo{journal}{New J. of Phys.} \textbf{\bibinfo{volume}{7}},
  \bibinfo{pages}{96} (\bibinfo{year}{2005}).

\bibitem[{\citenamefont{Zippilli and Morigi}(2005)}]{zippilli05}
\bibinfo{author}{\bibfnamefont{S.}~\bibnamefont{Zippilli}} \bibnamefont{and}
  \bibinfo{author}{\bibfnamefont{G.}~\bibnamefont{Morigi}},
  \bibinfo{journal}{quant-ph/0506030}  (\bibinfo{year}{2005}).

\bibitem[{\citenamefont{Vukics et~al.}(2005)\citenamefont{Vukics, Janszky, and
  Domokos}}]{vukics05}
\bibinfo{author}{\bibfnamefont{A.}~\bibnamefont{Vukics}},
  \bibinfo{author}{\bibfnamefont{J.}~\bibnamefont{Janszky}}, \bibnamefont{and}
  \bibinfo{author}{\bibfnamefont{P.}~\bibnamefont{Domokos}},
  \bibinfo{journal}{Journal of Physics B: At. Mol. Opt. Phys.}
  \textbf{\bibinfo{volume}{38}}, \bibinfo{pages}{1453} (\bibinfo{year}{2005}).

\end{thebibliography}

\end{document}